\documentclass[
 aip,
 amsmath,amssymb,
preprint,
]{revtex4-1}
\usepackage{color}
\usepackage{ulem}
\usepackage{amsmath}
\usepackage{subfigure}
\usepackage{graphicx}
\usepackage{dcolumn}
\usepackage{bm}

\usepackage[utf8]{inputenc}
\usepackage[T1]{fontenc}
\DeclareUnicodeCharacter{0306}{\={}}
\usepackage{mathptmx}
\usepackage{etoolbox}

\makeatletter
\def\@email#1#2{
 \endgroup
 \patchcmd{\titleblock@produce}
  {\frontmatter@RRAPformat}
  {\frontmatter@RRAPformat{\produce@RRAP{*#1\href{mailto:#2}{#2}}}\frontmatter@RRAPformat}
  {}{}
}
\makeatother
\begin{document}

\title{MHD simulation study on impurity assimilation efficiency and disruption dynamics during shattered pellet injection}

\author{Jinqiang Mao}
\affiliation{
	School of Physics, Huazhong University of Science and Technology, Wuhan, Hubei, 430074, China
}
\affiliation{
	State Key Laboratory of Advanced Electromagnetic Technology, International Joint Research Laboratory of Magnetic Confinement Fusion and Plasma Physics, School of Electrical and Electronic Engineering, Huazhong University of Science and Technology, Wuhan, Hubei, 430074, China
}
\author{Ping Zhu}
\email{zhup@hust.edu.cn} 
\affiliation{
	State Key Laboratory of Advanced Electromagnetic Technology, International Joint Research Laboratory of Magnetic Confinement Fusion and Plasma Physics, School of Electrical and Electronic Engineering, Huazhong University of Science and Technology, Wuhan, Hubei, 430074, China
}
\affiliation{
	Department of Nuclear Engineering and Engineering Physics, University of Wisconsin-Madison, Madison, WI 53706, United States of America
}

\author{Shiyong Zeng}
\affiliation{
	State Key Laboratory of Advanced Electromagnetic Technology, International Joint Research Laboratory of Magnetic Confinement Fusion and Plasma Physics, School of Electrical and Electronic Engineering, Huazhong University of Science and Technology, Wuhan, Hubei, 430074, China
}

\date{\today}

\begin{abstract}
Shattered Pellet Injection (SPI) has become a critical technique for mitigating plasma disruptions in fusion devices, yet optimizing its efficiency demands a proper understanding of the interaction between impurity dynamics and MHD response. We perform 3D nonlinear MHD simulations of SPI-induced disruption in a J-TEXT-like tokamak using the NIMROD code, systematically examining key parameters: fragment velocity and fineness, injection quantity, impurity composition, injection location and multiple injectors, resistivity, and parallel thermal conductivity. We find that slower fragment velocity enhances impurity assimilation and amplifies MHD activity. Finer fragments significantly increase impurity ablation and cooling efficiency. Mixed deuterium-neon pellets effectively elevate electron density without compromising radiative cooling efficiency. Plasma poloidal rotation affects ablation and cooling efficiency, whereas toroidally uniform multi-pellet injection enhances impurity ablation by nearly a factor equal to the number of pellets and lowers radiation asymmetry. Higher plasma parallel thermal conductivity results in higher radiation cooling efficiency in parallel directions, enhances impurity transport, and reduces toroidal peaking factor (TPF) of radiation. Variations in resistivity significantly influence Ohmic heating, impurity deposition and current dynamics after TQ, with higher resistivity leading to stronger magnetic perturbations and more pronounced current spikes. These findings provide physical bases for optimizing SPI schemes in future tokamak devices.
\end{abstract}

\maketitle

\section{Introduction}
Plasma disruptions pose a serious threat to the stable operation of tokamak fusion devices, primarily in three aspects: thermal loads, electromagnetic forces, and runaway currents\cite{lehnen_disruptions_2015}. To mitigate the harmful effects of disruptions, the Massive Gas Injection (MGI) scheme has been tested in experiments to dissipate plasma thermal energy and reduce mechanical stress from halo currents\cite{whyte_mitigation_2002}. However, due to the limited penetration depth of MGI often not far from the plasma edge, the effectiveness of MGI in disruption mitigation is significantly diminished in large tokamaks\cite{hollmann_measurements_2008}. Consequently, the Shattered Pellet Injection (SPI) scheme has been developed and eventually selected as the primary disruption mitigation system for ITER\cite{baylor_pellet_2009}. Comparisons between deuterium SPI and MGI experiments on DIII-D suggest SPI's superiority in penetrating the plasma\cite{commaux_demonstration_2010}. Subsequent experiments with neon SPI further show that with the same amount of neon, SPI can generate faster and stronger density perturbations and achieve higher impurity assimilation rates than MGI\cite{commaux_first_2016}. Following this, several tokamaks, including JET, KSTAR, J-TEXT, HL-2A, and EAST, have also developed SPI systems\cite{sheikh_disruption_2021,park_deployment_2020,li_design_2018,li_comparison_2021,xu_preliminary_2020,Yuan_2023}.

Experiments on DIII-D demonstrate that SPI with mixed species offers a possible approach to tuning disruption mitigation\cite{shiraki_thermal_2016}. Complementary studies on JET investigated the effect of Neon content in deuterium-neon SPI mixtures, revealing that higher impurity concentrations lead to shorter thermal quench (TQ) and current quench (CQ) durations\cite{sheikh_disruption_2021}. The subsequent SPI experiments on DIII-D have investigated the dual shattered pellet injection\cite{herfindal_injection_2019}, the penetration in low and high energy plasma\cite{raman_shattered_2020}, the 3D radiation, density, and MHD structure\cite{sweeney_3d_2021}, and the comparisons between low-Z and high-Z SPI\cite{lvovskiy_density_2023}. JET has conducted a detailed evaluation of the mitigation efficiency of SPI\cite{jachmich_shattered_2021}. Experiments on the Large Helical Device (LHD) find that the addition of a small amount of Neon to mixed pellets ($H_{2}+Ne$) injected on the low-field side results in deeper penetration. This is primarily due to the strong radiation from Neon, which suppresses the formation of dense plasmoids that hinders the penetration of the pellets and reduces the $\vec{E} \times \vec{B}$ drift of ablated material\cite{matsuyama_enhanced_2022}.

Previous NIMROD simulations indicate that mixed pellets can lead to more benign disruptions\cite{kim_shattered_2019}. MHD simulations of SPI on JET using M3D-C1 and NIMROD find that reducing plasma viscosity increases MHD activity and decreases thermal quench time slightly. The simulation results are found qualitatively similar for those cases with a single monolithic pellet, a pencil beam plume, or the realistic plume\cite{mcclenaghan_mhd_2023}. Additionally, the JOREK simulation of SPI on JET shows that the convective transport caused by core MHD instabilities (e.g. 1/1 kink) contributes significantly to the penetration of impurities into the core\cite{hu_3d_2018}. The simulation study of deuterium SPI on ASDEX Upgrade investigated how MHD activity affects shard ablation\cite{hoelzl_first_2020}. Simulations of pure deuterium SPI in ITER find the possibility of diluting the plasma by more than a factor 10 without triggering large MHD activity\cite{nardon_fast_2020}. Simulation study using JOREK investigates radiation asymmetry and MHD responses, concluding that  in scenarios of strong radiative cooling ($Ar$ SPI), the contraction of the axisymmetric component of the current profile is the primary driver of MHD instability. In contrast, under weaker radiative cooling conditions ($D_{2}$ SPI), helical cooling on each major rational surface dominates. The triggering of TQ and the transport of impurities are closely linked to the distribution of the safety factor profile and the structure of MHD modes. Toroidally symmetric dual SPI injection effectively reduces the toroidal radiation asymmetry observed in a single SPI\cite{hu_radiation_2021}. Simulations have examined how the injection position of SPI relative to pre-existing 2/1 magnetic islands affects subsequent island growth and the progression of TQ\cite{wieschollek_role_2022}. The JOREK code has later included a non-equilibrium collisional-radiative impurity model, which has been compared with the previously used coronal equilibrium model\cite{hu_collisional-radiative_2021}. The effects of the pellet ablation models, the toroidal density distribution, the background impurity radiation, the parallel heat transport models\cite{hu_collisional-radiative_2023}, and the hot-tail electrons on assimilation of impurities\cite{hu_hot-tail_2022}, as well as the plasmoid drift and rocket effects on impurity penetration and plasma response are also investigated using JOREK\cite{kong_interpretative_2024}.

Despite these studies, there remains no consensus regarding how to regulate the disruption mitigation efficiency or enhance the assimilation of impurities, or how SPI and plasma parameters may influence on the disruption mitigation efficiency. To address these issues, in this work the SPI disruption mitigation process in a J-TEXT-like tokamak plasma is simulated and analyzed using the NIMROD code\cite{glasser_nimrod_1999,sovinec_nonlinear_2004}, which incorporates an atomic and radiation physics model adapted from KPRAD\cite{whyte_energy_1997} and a particle based shattered pellet injection model\cite{kim_shattered_2019}. We then perform a parameter scan to systematically assess the influence of key variables on both disruption dynamics and impurity assimilation. The impurity assimilation efficiency is determined in this work in terms of the total quantity, spatial distribution, and temporal evolution characteristics of impurity atoms that are successfully ionized and ultimately retained within the plasma volume during nearly the entire disruption process -- from the initial contact of impurity fragments with the plasma through to the plasma CQ phase.

The rest of this paper is organized as follows: Section 2 describes the simulation model and setup. Section 3 presents the general results of plasma rapid shutdown triggered by SPI. Section 4 demonstrates the effects of key SPI and plasma parameters on the disruption characteristics. And finally, section 5 gives the discussion and conclusion.

\section{Model and simulation setup}

\subsection{NIMROD/KPRAD model}
The simulations in this work are based on the single-fluid resistive MHD model implemented in the NIMROD code\cite{glasser_nimrod_1999,sovinec_nonlinear_2004}. The calculation of impurity ionization and radiation is based on the KPRAD model\cite{whyte_energy_1997}. In the single-fluid model, all species share the same temperature and fluid velocity, which assumes instant thermal equilibration among main ion, electron and impurity species. The number densities of each ionization state for the main ions and impurities advance independently. The equations for the impurity MHD model are shown as follows:

\begin{equation}
	\rho \frac{\mathrm{d} \vec{V}}{\mathrm{~d} t}=-\nabla p+\vec{J} \times \vec{B}+\nabla \cdot(\rho \nu \nabla \vec{V}) \label{eq1}
\end{equation}
\begin{equation}
	\frac{\mathrm{d} n_{i}}{\mathrm{~d} t}+n_{\mathrm{i}} \nabla \cdot \vec{V}=\nabla \cdot\left(D \nabla n_{\mathrm{i}}\right)+S_{{ion } / 3-{ body }} \label{eq2}
\end{equation}
\begin{equation}
	\frac{\mathrm{d} n_{Z, Z=0-10}}{\mathrm{~d} t}+n_{\mathrm{Z}} \nabla \cdot \vec{V}=\nabla \cdot\left(D \nabla n_{Z}\right)+S_{{ion} / \mathrm{rec}} \label{eq3}
\end{equation}
\begin{equation}
	n_{\mathrm{e}} \frac{\mathrm{d} T_{\mathrm{e}}}{\mathrm{d} t}=(\gamma-1)\left[n_{\mathrm{e}} T_{\mathrm{e}} \nabla \cdot \vec{V}+\nabla \cdot \overrightarrow{q_{\mathrm{e}}}-Q_{{loss }}\right] \label{eq4}
\end{equation}
\begin{equation}
	\vec{q}_{\mathrm{e}}=-n_{\mathrm{e}}\left[\kappa_{\|} \hat{b} \hat{b}+\kappa_{\perp}(\mathcal{I}-\hat{b} \hat{b})\right] \cdot \nabla T_{\mathrm{e}} \label{eq5}
\end{equation}
\begin{equation}
	\vec{E}+\vec{V} \times \vec{B}=\eta \vec{J} \label{eq6}
\end{equation}

\noindent$n_{i}$, $n_{e}$, $n_{Z}$ are the main ion, electron, and impurity ion number density respectively, $\rho$, $\vec{V}$, $\vec{J}$, and $p$ the plasma mass density, velocity, current density and pressure. $T_{e}$ and $\vec{q_{e}}$ are the electron temperature and heat flux respectively. $D$, $\nu$, $\eta$ and $\kappa_{\parallel}$ ($\kappa_{\perp}$) are the plasma diffusivity, kinematic viscosity, resistivity and parallel (perpendicular) thermal conductivity. $\gamma$ is the adiabatic index. $S_{{ion} / \mathrm{rec}}$ is the density source from ionization and recombination, and $S_{{ion } / 3-{ body }}$ also includes contribution from three-body recombination, $\hat{b}=\vec{B}/B$ and $\mathcal{I}$ is the unit dyadic tensor. Pressure $p$ and mass density $\rho$ in momentum equation (1) include impurity contributions. The power loss term $Q_{loss }$ in equation (4) is calculated from the KPRAD module based on a coronal model and includes contributions from the Ohmic heating and the losses due to ionization, recombination and radiation. 

\subsection{Particle based shattered pellet injection model}
The NIMROD code has implemented a particle-based SPI model that considers fragments as discrete particles, assumes the fragment number small without interaction, and neglects the fragment geometry and details of ablation cloud\cite{kim_shattered_2019}. The ablation of the fragments is calculated using an ablation function $G\left ( n_{e},T_{e},r_{f},X \right ) \propto \lambda \left ( X \right ) n_{e}^{1 / 3} T_{e}^{5 / 3} r_{f}^{4 / 3}$, which depends on the electron density, electron temperature, fragment radius and composition\cite{parks_theoretical_2017}. The parameter $\lambda$ is a function of the mixture ratio $ X = {m_{D_{2}}} / (m_{Ne} + m_{D_{2}}) $, which is the molar fraction of deuterium in the pellet. The electron density and temperature for the pellet are computed by averaging the local background plasma values within an estimated radius $r_{e}$ . The deposition of impurities in the finite element poloidal plane is modeled using a Gaussian circle with radius $\Delta r_{d}$, whereas the deposition in the toroidal direction is modeled using a Gaussian profile with $\Delta \phi_{d}$. The ablated material exists initially as neutral gas, and is considered assimilated into the plasma upon deposition and subsequent ionization.

In the SPI module, we can set independent initial position, initial velocity, initial size, sedimentation radius, evaluation radius, deposition offset and other parameters for each particle, which corresponds to N fragments with the same properties. Through adjusting input parameters, various shapes of plume can be specified, such as the single particle plume, the pencil beam plume and the particle cloud plume, for examples\cite{kim_shattered_2019}.

\subsection{Simulation setup}
\begin{figure}[htbp]
	\centering
	\includegraphics[width=\linewidth]{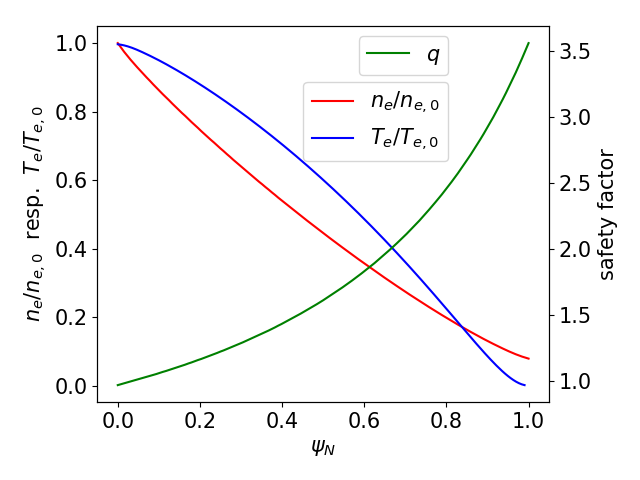}
	\caption{Initial equilibrium profiles as functions of the normalized poloidal flux. The temperature profile is shown in blue and the density profile in red, both normalized to their core values (left y-axis). The safety factor profile is shown in green (right y-axis).}
	\label{fig:1}
\end{figure}

The initial equilibrium is based on a J-TEXT-like circular-shaped limiter tokamak plasma\cite{zeng_mhd_2021,ding_overview_2024}. The initial flux-surface-averaged $T_{e}$, $n_{e}$, and $q$ profiles are shown in Fig. \ref{fig:1}. The initial plasma is stationary, $V_{0}=0m/s$. The Spitzer resistivity and the Braginskii thermal conductivity models are adopted. We limited the maximum resistivity to avoid both excessive Ohmic heating at low temperatures and potential numerical convergence issues. The Lundquist number on axis is set to $10^7$. Anisotropic thermal conductivities are temperature dependent, $\kappa_{\|}=\kappa_{\| 0}\left(T_{e} / 700 \mathrm{eV}\right)^{5 / 2} \mathrm{~m}^{2} / \mathrm{s}$ and $\kappa_{\perp}=\kappa_{\perp 0}\left(700 e V / T_{e}\right)^{1 / 2}\left(1 / B^{2}\right) m^{2} / \mathrm{s}$. The key parameters are listed in Table~\ref{tab:table1}.

\begin{table*}
	\caption{\label{tab:table1}Key parameters in the simulation.}
	\begin{ruledtabular}
		\begin{tabular}{lccc}
			Parameter & Symbol & Value & Unit \\
			\hline
			Minor radius & a & 0.25 & $m$ \\ 
			Major radius & $R_{0}$ & 1.05 & $m$ \\
			Plasma current & $I_{p}$ & 150 & $kA$ \\
			Toroidal magnetic field & $B_{t0}$ & 1.74 & $T$ \\
			Core value of safety factor & $q_{0}$ & 0.95 & Dimensionless \\
			Edge value of safety factor & $q_{a}$ & 3.56 & Dimensionless \\
			Core electron density & $n_{e0}$ & $1.875\times 10^{19}$ & $m^{-3}$ \\
			Core electron temperature & $T_{e0}$ & 700 & $eV$ \\
			Equilibrium velocity & $V_{0}$ & 0 & $m/s$ \\
			The core resistivity & $\eta_{0}$& $4.3\times 10^{-8}$ & $\Omega m$ \\
			Kinematic viscosity & $\nu$ & 27 & $m^{2}/s$ \\
			The core Lundquist number & $S_{0}$ & $1\times 10^{7}$ & Dimensionless \\
			Core perpendicular thermal conductivity & $\kappa_{\perp0}$ & 1 & $m^{2}/s$ \\
			Core parallel thermal conductivity & $\kappa_{\parallel0}$ & $10^{6}$ & $m^{2}/s$ \\
			Diffusivity & $D$ & 2 & $m^{2}/s$ \\
			Radius of evaluation & $r_{e}$ & 0.04 & $m$ \\
			Poloidal deposition & $\Delta r_{d}$ & 0.04 & $m$ \\
			Toroidal deposition & $\Delta \phi_{d}$ & $0.4\pi$ & rad \\
		\end{tabular}
	\end{ruledtabular}
\end{table*}

\begin{figure}[htbp]
	\centering
	\includegraphics[width=\linewidth]{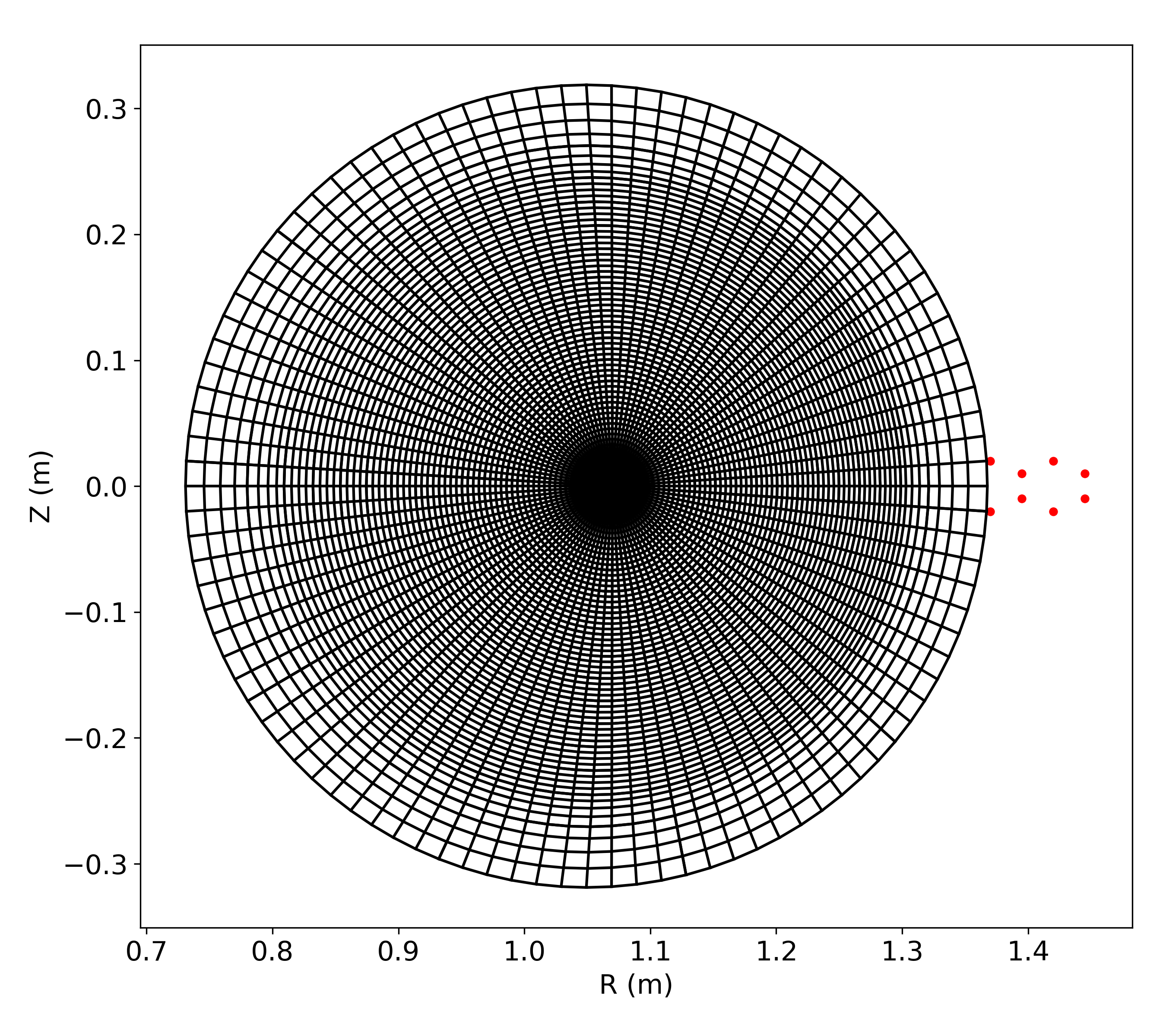}
	\caption{Simulation mesh grid and initial fragment positions (red dots).}
	\label{fig:2}
\end{figure}

The poloidal grid includes $64\times 100$ 2D bicubic finite elements, along with 6 toroidal Fourier modes n = [0, 5]. Here n is the toroidal mode number. The grid resolution is sufficient to capture and reproduce most of the characteristic events observed in SPI disruption experiments. The simulation mesh grid and initial fragment positions are shown in Fig. \ref{fig:2}.

The setting of the fragment plume is 8 small fragments with a radius of 0.5mm, containing a total of $ 1.8\times 10^{20}$ Ne atoms. All the fragments travels at a speed of 240 m/s. The trajectory originates from the vacuum region on the low-field side of the equatorial plane and extends towards the plasma core. The plume initially measures 10 cm in length and 4 cm in width. The spread angle is 10 degree. The choice of deposition parameters may significantly influence the simulation outcomes. An excessively large $\Delta r_{d}$ may result in a loss of detail in plasma profile evolution, leading to substantial deviations from realistic conditions. Conversely, an overly small $\Delta r_{d}$ can cause locally concentrated impurity deposition and strong perturbations, introducing numerical instability. In simulations, increasing $\Delta \phi_{d}$ helps mitigate toroidal asymmetry, while overly small values may again lead to numerical issues due to localized deposition accumulation.

\section{Characteristic SPI disruption}
\begin{figure}[htbp]
	\centering
	\includegraphics[width=\linewidth]{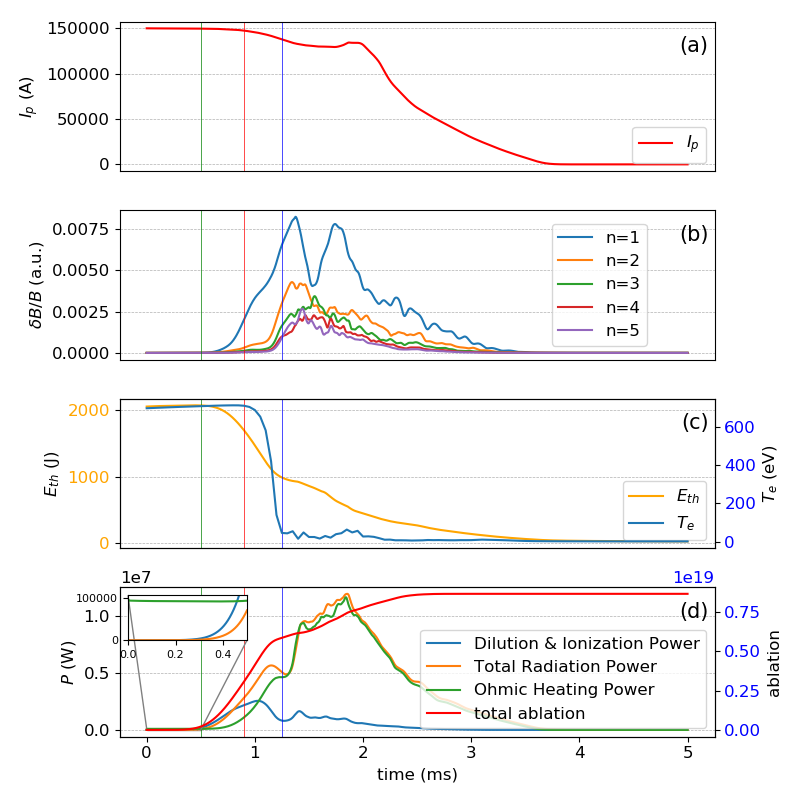}
	\caption{(a) Plasma current, (b) normalized magnetic energies of toroidal components $\delta B/B = \sqrt{\left ( W_{mag,n} / W_{mag,n=0} \right ) }$, (c) core electron temperature at $\phi=0$ (the solid blue line) and thermal energy (the solid yellow line), and (d) sum of dilution power and ionization power (the solid blue line), Ohmic heating power (the solid green line) and total radiation power (the solid yellow line) as functions of time. The green, red, and blue vertical lines correspond to 0.5 ms, 0.9 ms, and 1.25 ms, respectively.}
	\label{fig:3}
\end{figure}

The evolution of some key plasma parameters during a $Ne$ SPI process is illustrated in Fig. \ref{fig:3}. During the impurity flight phase (0 - 0.5 ms), impurities are injected from the low-field side of the equatorial plane and propagate towards the plasma boundary. Before the impurities reach the boundary, the plasma internal energy, the core temperature, and the plasma current all remain unchanged. Following the impurity flight phase, the pre-TQ phase (0.5 - 0.9 ms) begins when the impurities reach the plasma boundary. During this phase, the impurities gradually penetrate towards the plasma core, leading to the cooling of plasma boundary. This cooling results in a gradual decay of the internal energy, with approximately 15\% of the thermal energy being dissipated by the end of pre-TQ phase. However, since the impurities have not yet reached the core, the core temperature and the total plasma current continue to experience minimal variation. As the impurity cold front penetrates deeper, the perturbations induced by impurity radiation cooling intensify, with the \( n = 1 \) mode dominating throughout the process [Fig. \ref{fig:3}(b)]. At this stage, the Ohmic heating power exceeds the power loss from impurity dilution, ionization, and radiation (as shown in the inset of Fig. \ref{fig:3}(d)), resulting in a temporary increase in plasma internal energy. As the pre-TQ phase progresses, impurity radiation and magnetic perturbations grow, setting the stage for the subsequent TQ phase, which begins with the collapse of the plasma core temperature and ends with a decrease of over 90\% in the core temperature, lasting approximately 0.35 ms from 0.9 ms to 1.25 ms. It is worth noting that the core temperature is taken as the temperature at the toroidal position same as the injection location with $ \phi=0 $. During the TQ phase, the plasma core temperature collapses, and completely cools down to a level of several tens of $eV$, when the amplitude of magnetic perturbations has also increased. By the end of TQ phase, approximately 50\% of the thermal energy has been dissipated, with the remaining thermal energy residing in the high-field side and other residual hot regions of the plasma in toroidal direction. After 1.25 ms, the amplitude of magnetic perturbation reaches its peak, and a current spike becomes prominent right before the start of CQ phase. Simultaneously, the Ohmic heating power increases and approximately balances the radiative power from the impurities [Fig. \ref{fig:3}(d)]. The magnetic energy of the plasma, which was barely dissipated during the TQ phase, begins to radiate and dissipate through Ohmic power during the CQ phase until the plasma discharge terminates. The current spike and the peak in magnetic perturbations are key indicators of the transition into the CQ phase, highlighting the complex interplay between magnetic fields, plasma current, and impurity radiation cooling during a plasma disruption event.

\begin{figure}[htbp]
	\centering
	\includegraphics[width=0.8\linewidth]{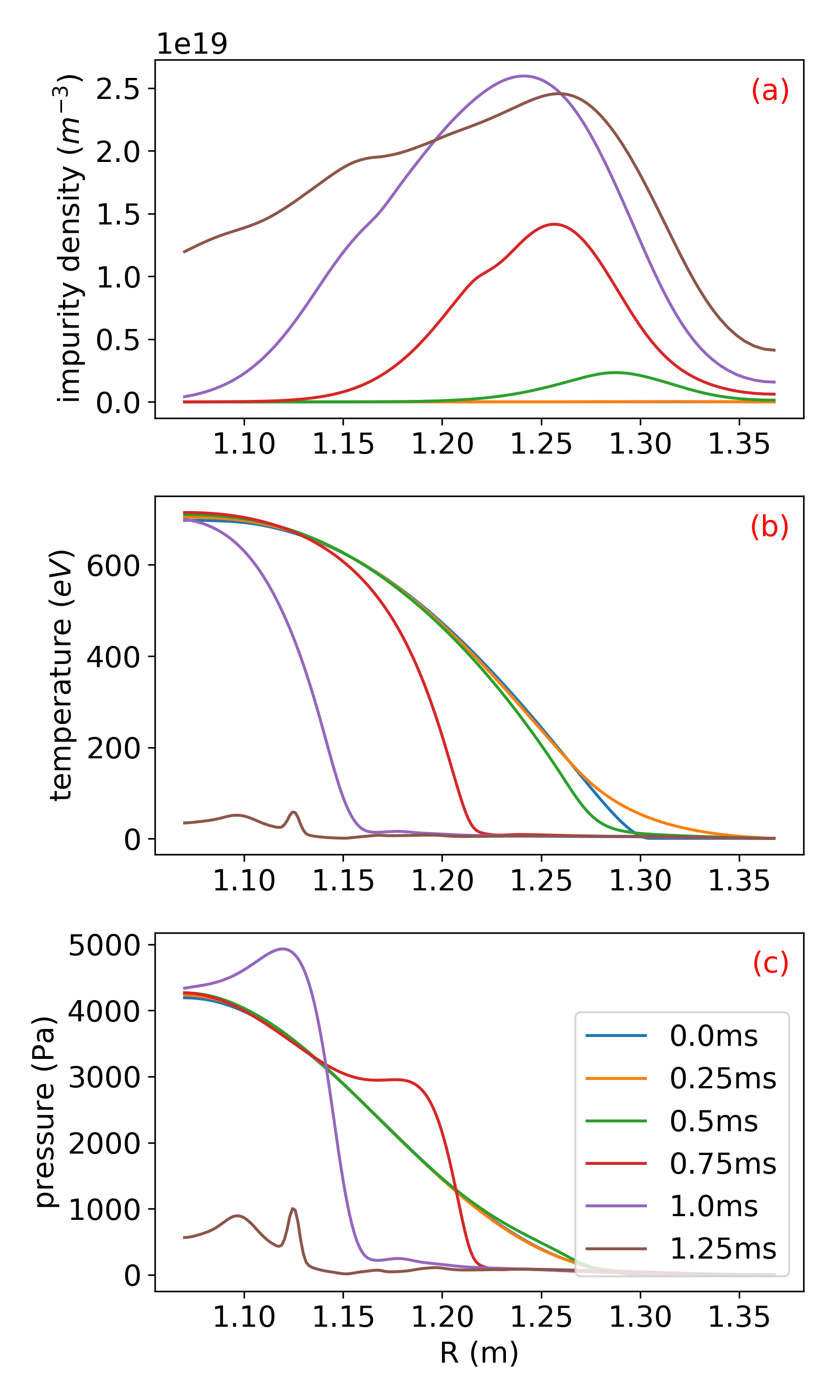}
	\caption{The time evolution of (a) impurity deposition density, (b) temperature, and (c) pressure profiles as functions of major radius.}
	\label{fig:4}
\end{figure}

\begin{figure}[htbp]
	\centering
	\includegraphics[width=\linewidth]{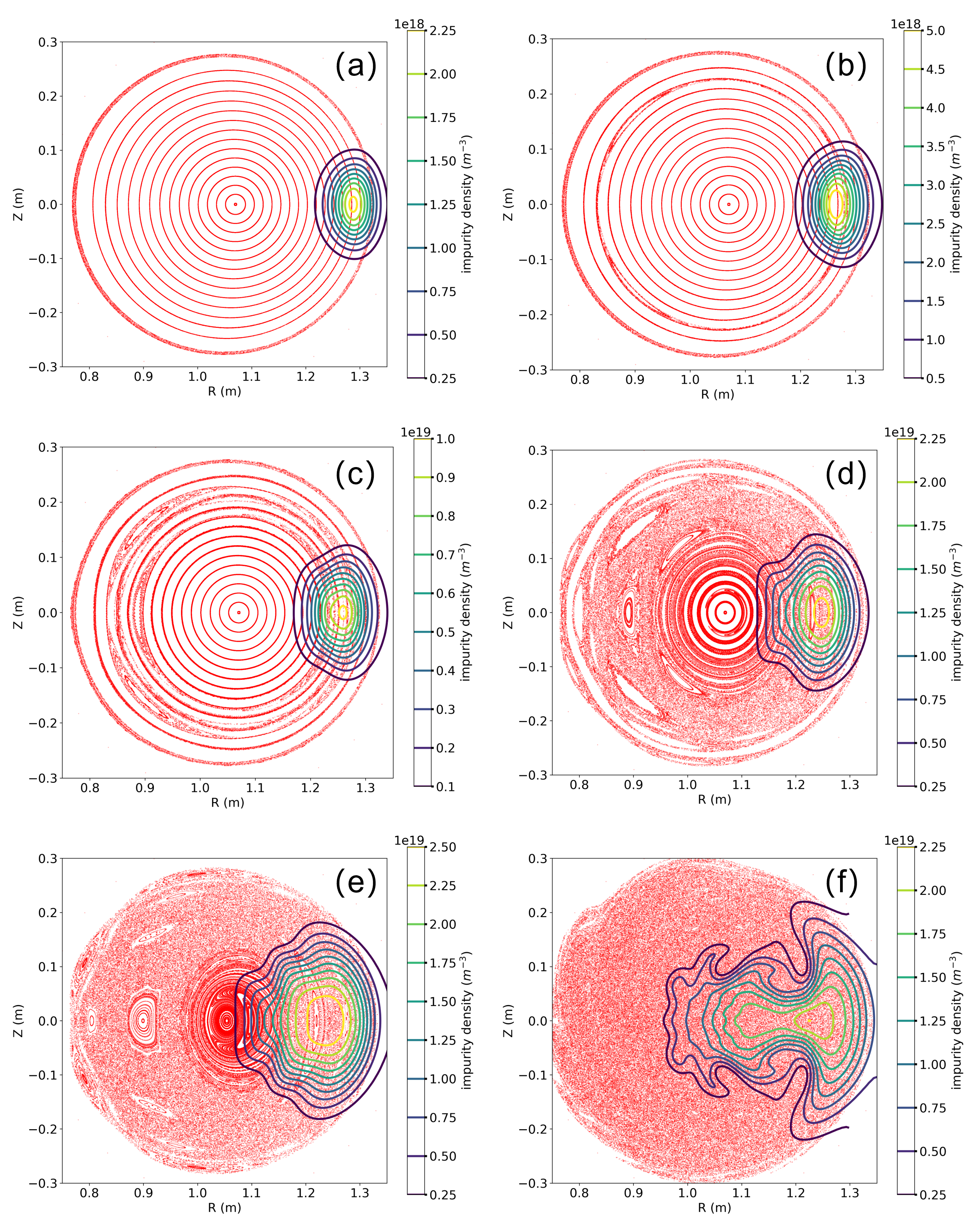}
	\caption{Poincaré plots and impurity deposition distribution in poloidal plane at various times: (a) t=0.5ms, (b) t=0.6ms, (c) t=0.7ms, (d) t=0.9ms, (e) t=1.1ms, and (f) t=1.4ms.}
	\label{fig:5}
\end{figure}

\begin{figure}[htbp]
	\centering
	\includegraphics[width=\linewidth]{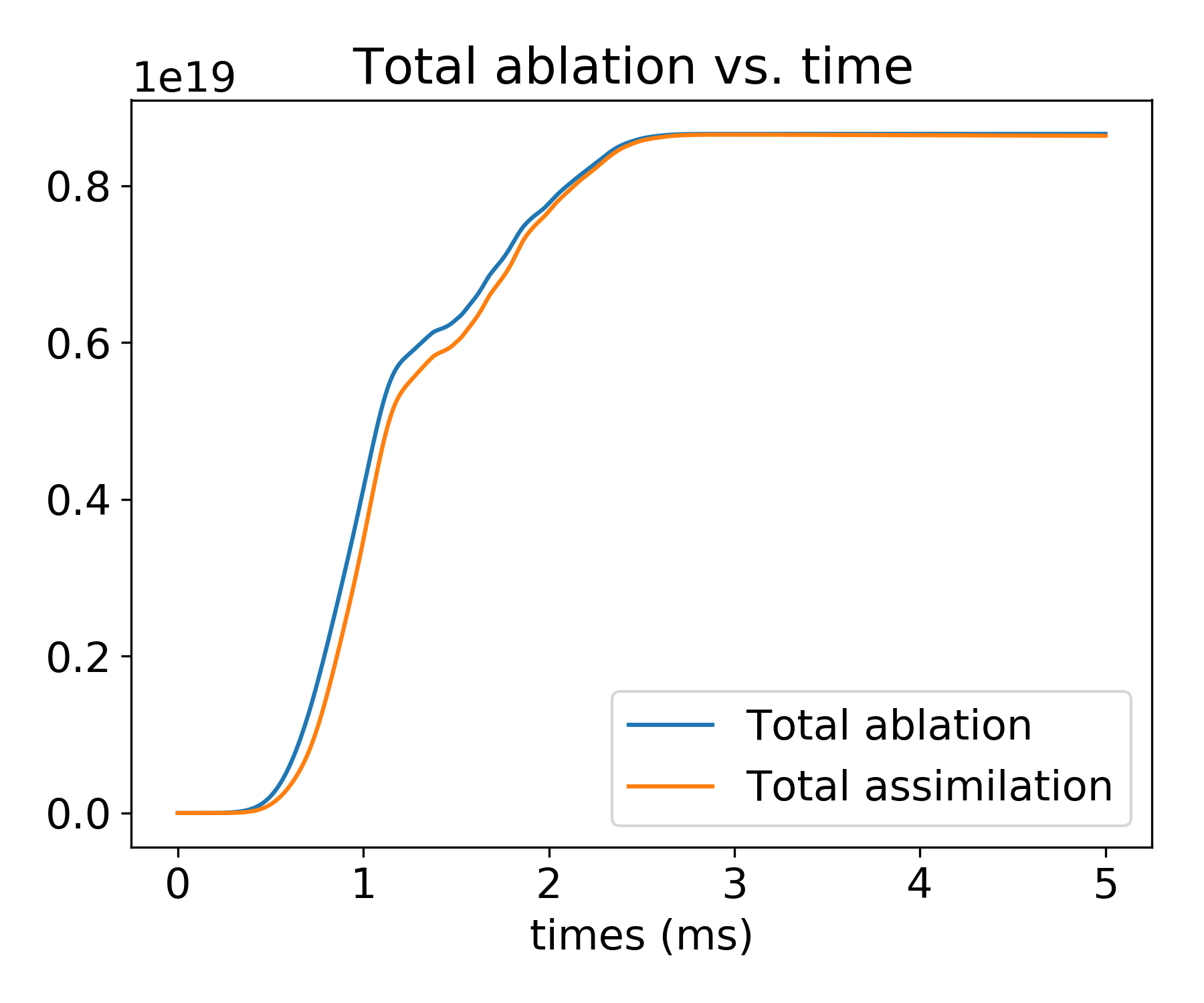}
	\caption{Total ablation and assimilation of all the fragments as functions of time.}
	\label{fig:6}
\end{figure}

The radial profile of the impurity density, temperature, and pressure profiles are plotted along the chord extending from the magnetic axis to the SPI port location, which corresponds to the trajectory of the fragments [Fig. \ref{fig:4}]. As the impurities approach the core, the deposition rate increases due to higher temperature, which leads to increasingly significant changes in the pressure profile and the formation of localized bulges that drive strong flow and enhance parallel transportation\cite{kim_shattered_2019}. The Poincaré plot overlaid with the distribution of impurity deposition density illustrates the growth of magnetic islands during impurity penetration [Fig. \ref{fig:5}]. During the pre-TQ phase, impurities gradually penetrate the plasma [Fig. \ref{fig:5}(a)-(c)]. MHD modes begin to grow gradually as the impurities deposit. The 3/1 mode appears first, followed by the 2/1 mode. The moment just before the onset of TQ, the majority of the impurity deposition is concentrated on the \( q = 2 \) rational surface while reaching the \( q = 1 \) rational surface [Fig. \ref{fig:5}(d)]. During the TQ phase, the 2/1 magnetic island significantly enlarges and the 1/1 kink mode appears as the impurities penetrate and begin to cool down the plasma core, when most of the magnetic surfaces start to become stochastic [Fig. \ref{fig:5}(e)]. By the late TQ phase, the plasma is almost completely cooled down and the magnetic surfaces are fully stochastic [Fig. \ref{fig:5}(f)]. The peak impurity deposition is observed on the low-field side, which is due to the fact that the fragments are injected from the low-field side and no poloidal rotation occurs. Intense ablation occurs during 0.6-1.1ms, and almost all the ablated impurities are assimilated in the later stage [Fig. \ref{fig:6}]. The ablation rate of the impurity pellet is dependent on the background temperature, being proportional to the temperature raised to the power of 5/3. By the time the fragments reach the plasma core, the plasma temperature has significantly decreased, leading to a substantial reduction in the ablation rate, preventing deeper deposition. In fact, fragments used in this simulation do not completely ablate before the end of the disruption.

\section{Effect of different parameters on SPI}
To find out the key pellet and plasma effects on impurity assimilation and disruption dynamics during SPI, we have conducted comparative simulation analyses across several parameters including fragment velocity, fragment fineness, injection quantity, impurity composition, injection location and multiple injectors, plasma resistivity, and parallel thermal conductivity.

\subsection{Dependence on fragment velocity}
\begin{figure}[htbp]
	\centering
	\includegraphics[width=\linewidth]{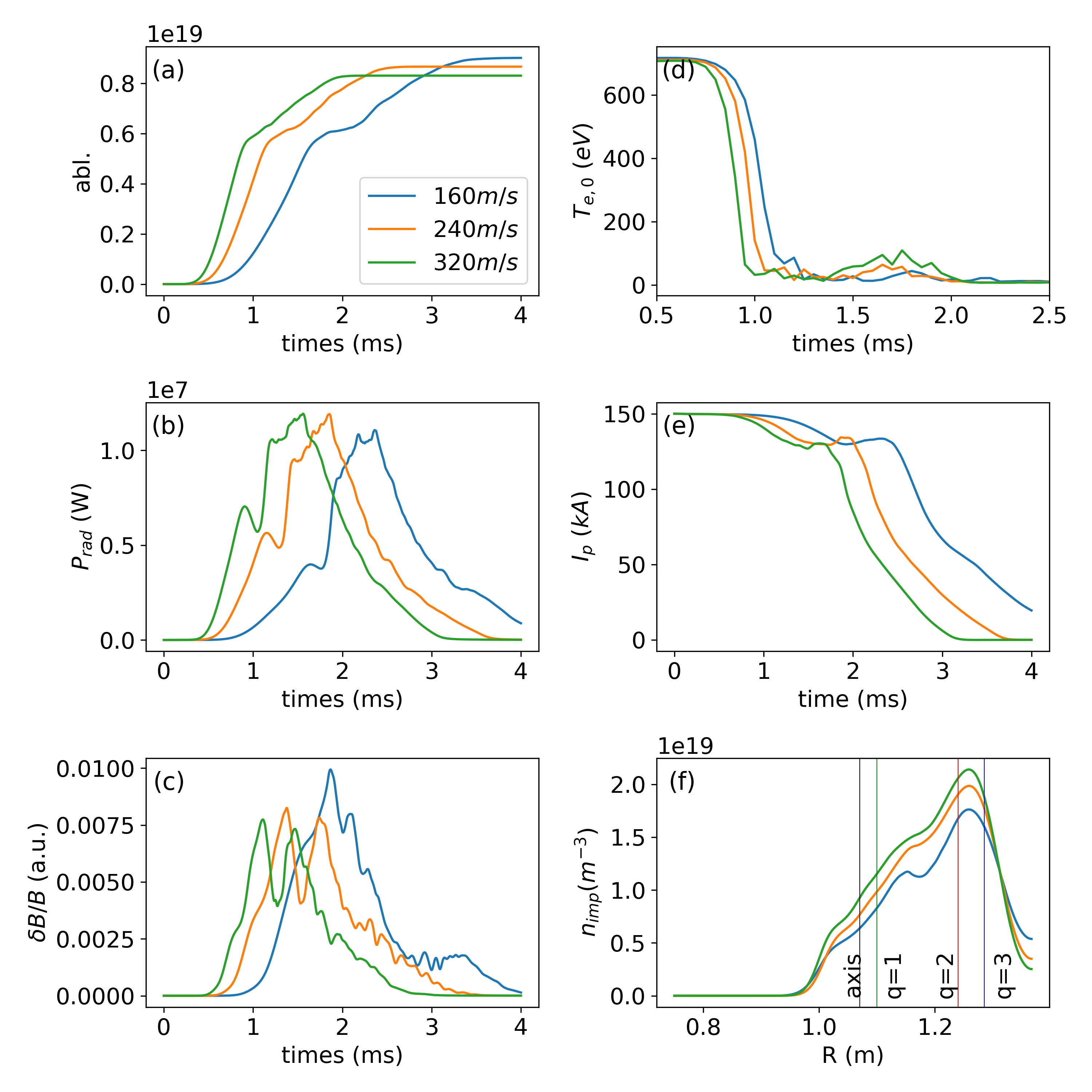}
	\caption{(a) Impurity ablation, (b) radiation power, (c) normalized magnetic energy of $n=1$ toroidal component $\delta B/B = \sqrt{\left ( W_{mag,n=1} / W_{mag,n=0} \right ) }$, (d) core plasma temperature at $\phi=0$ (shifted time relative to the start of TQ), and (e) plasma current as functions of time, and (f) impurity density profiles immediately after TQ as functions of major radius, for various initial pellet injection speeds. The black, green, red, and blue solid vertical lines represent the locations of axis, $q=1$, $q=2$, and $q=3$ rational surfaces respectively.}
	\label{fig:7}
\end{figure}

The fragment velocity is expected to have a major influence on radiation intensity ,impurity assimilation rate, TQ duration, magnetic perturbation intensity, and impurity penetration depth during the SPI disruption process. Three initial fragment velocity 160 m/s, 240 m/s, and 320 m/s are selected for comparison, which are all within typical experimental range. Experimental measurements from the Oak Ridge National Laboratory have found that there is a strong inverse relationship between pellet velocity and the size of the fragments, and the remaining solid mass decreases as velocity increases\cite{gebhart_experimental_2020,gebhart_shatter_2020}. This will inevitably affect the SPI process in experiment, but in our simulation, we do not consider the effect of velocity on fragment size in order to isolate the fragment speed effect itself. Simulations indicate that as the initial fragment velocity increases, the total impurity ablation decreases [Fig. \ref{fig:7}(a)], the peak value of radiation intensity slightly increases [Fig. \ref{fig:7}(b)], and the maximum of $ n = 1 $ component of magnetic perturbation decreases [Fig. \ref{fig:7}(c)]. Increasing velocity also slightly reduces the TQ duration [Fig. \ref{fig:7}(d), where the time is aligned with the TQ start time]. From Fig. \ref{fig:7}(f), it can be seen that the flight velocity has a significant impact on the impurity deposition depth and quantity. Higher velocity results in more impurities being deposited closer to the plasma core, leading to a higher impurity density profile in the mid-plane. Moreover, as illustrated in Fig. \ref{fig:7}(d), the TQ induced by higher initial fragment speed is less thorough or complete. The radial transport of impurities in the tokamak plasma is relatively slow, whereas the fragment velocity directly affects the duration of the pre-TQ and TQ stages. To accommodate these timescales, the fragment velocity required for the SPI system in large tokamaks should be relatively higher than those used in current simulations that are more intended for the medium-sized tokamaks such as J-TEXT.

\subsection{Dependence on fragment fineness}
\begin{figure}[htbp]
	\centering
	\includegraphics[width=\linewidth]{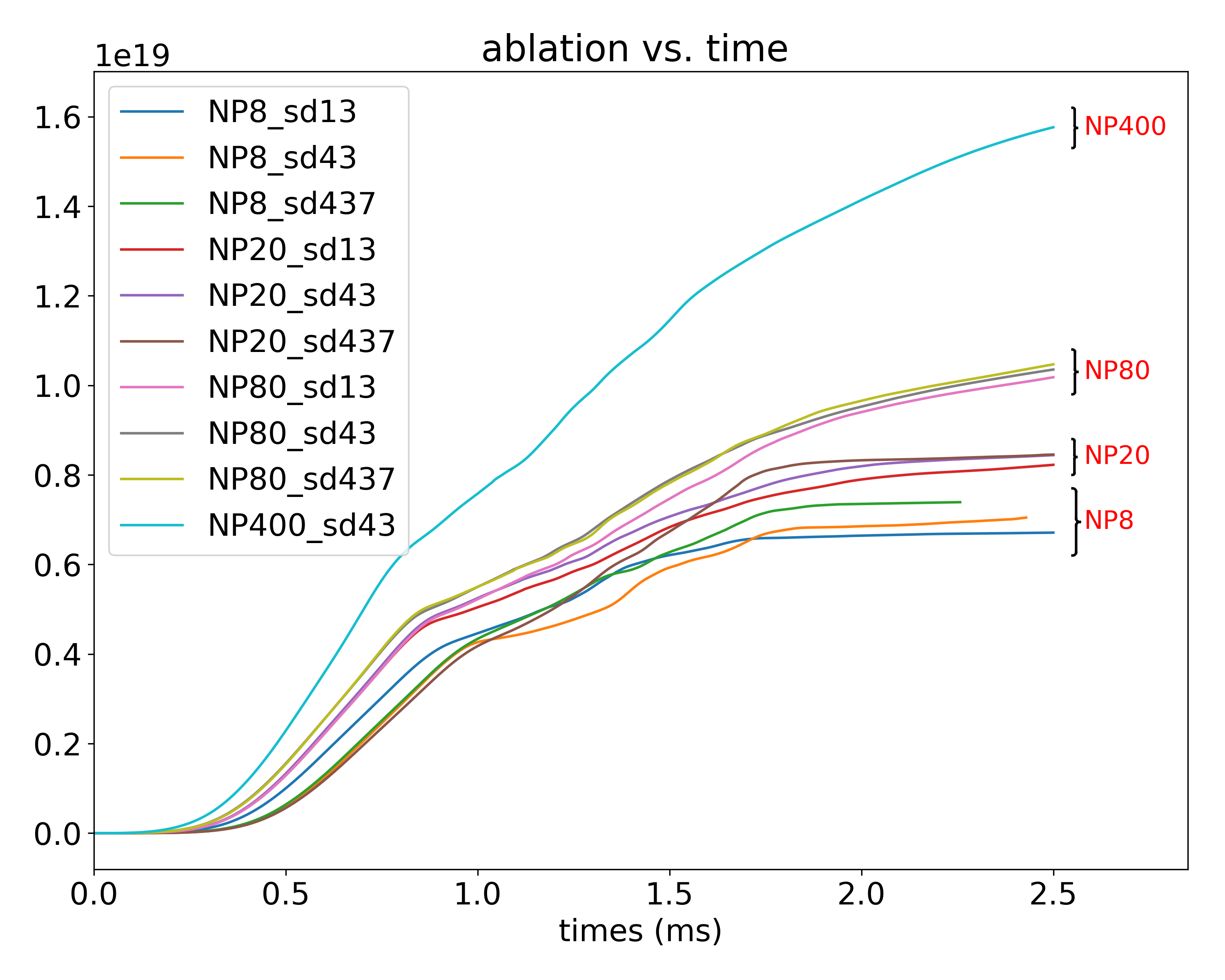}
	\caption{Impurity assimilations as functions of time for various fragment fineness and plume shapes. "NP\#" denotes the number of fragments, and "sd\#" denotes the type of plume shape.}
	\label{fig:8}
\end{figure}

\begin{figure}[htbp]
	\centering
	\includegraphics[width=0.8\linewidth]{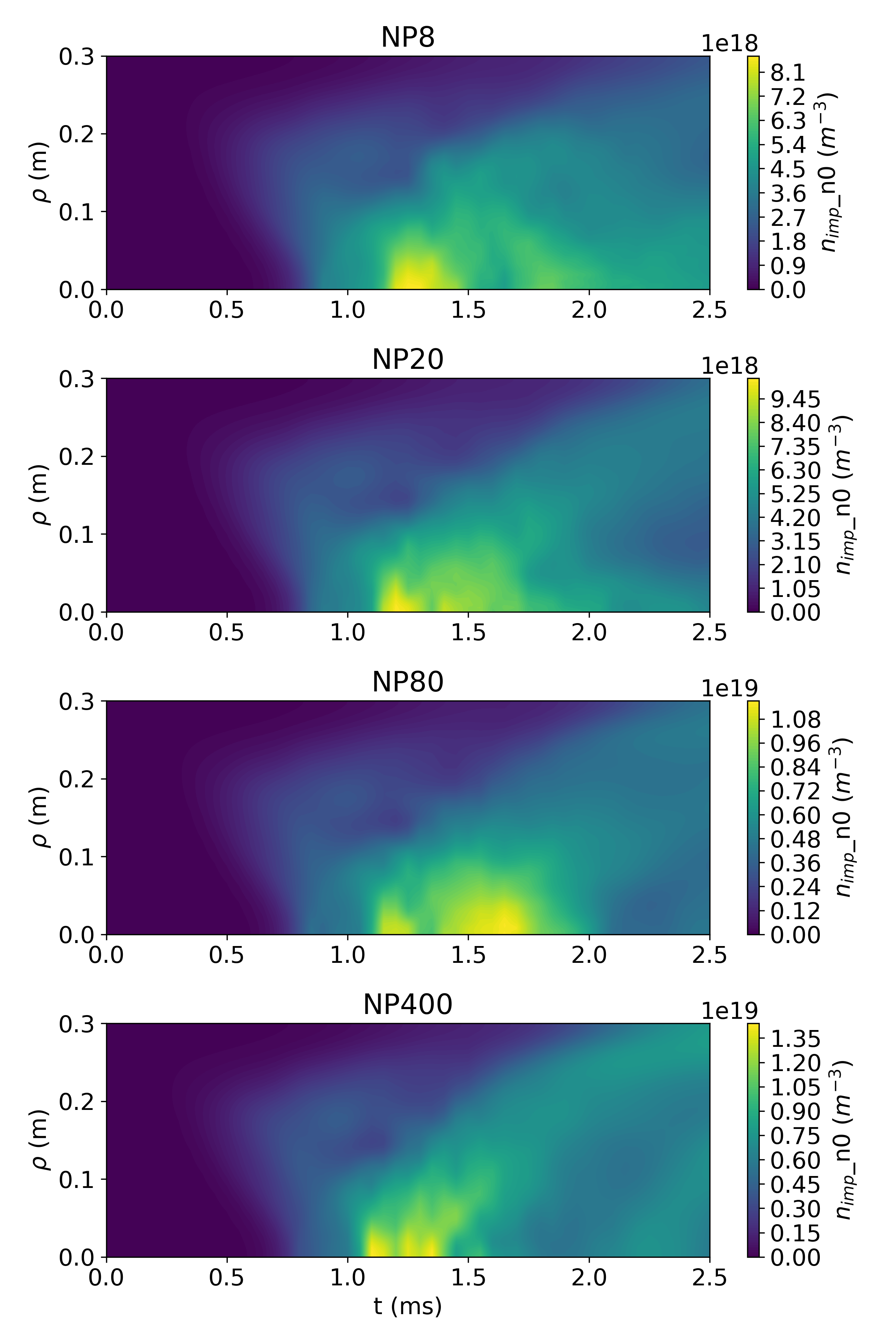}
	\caption{Radial distributions of impurity density ($n=0$ Fourier component) as functions of time for various fragment numbers.}
	\label{fig:9}
\end{figure}

\begin{figure}[htbp]
	\centering
	\includegraphics[width=0.8\linewidth]{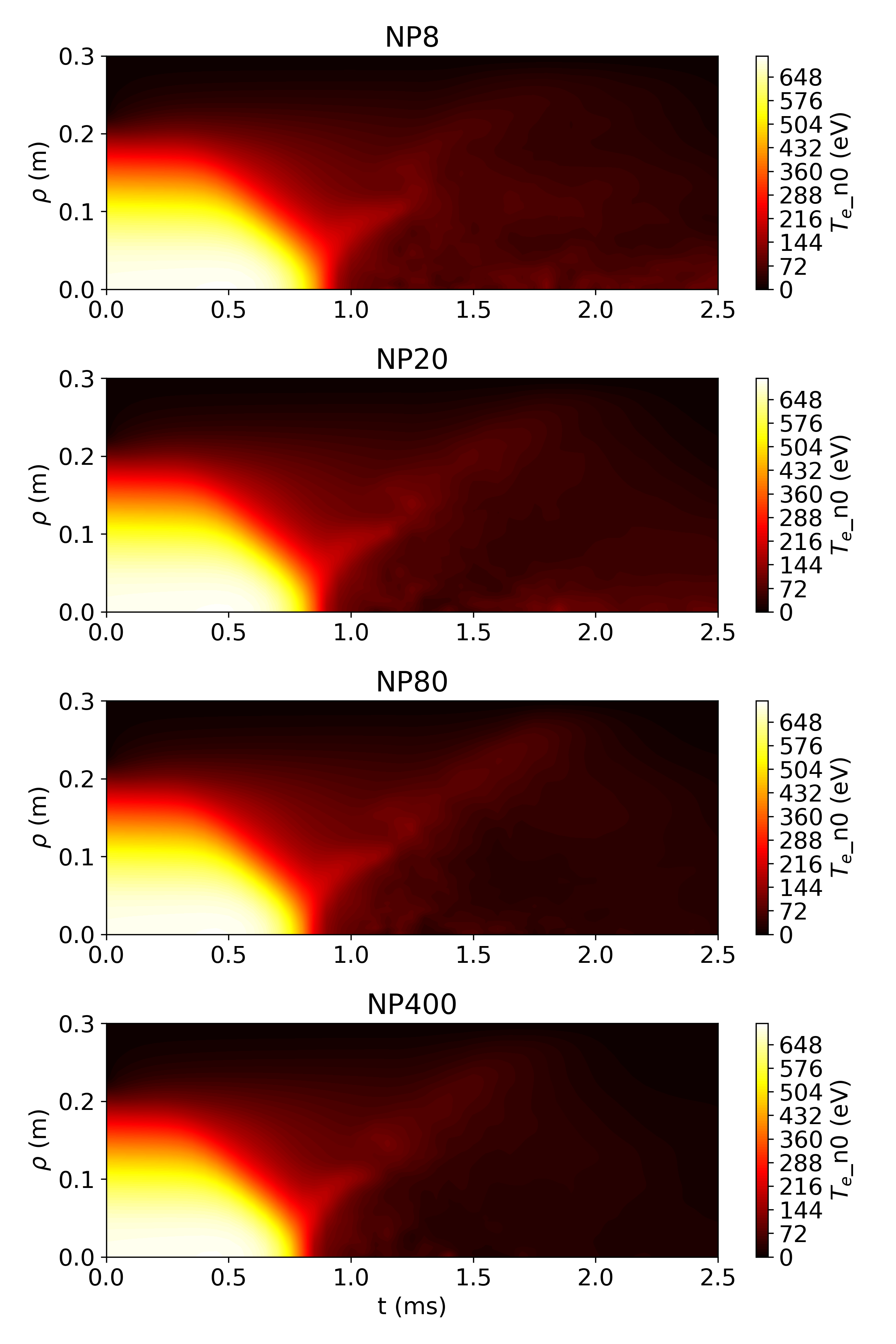}
	\caption{Radial distributions of plasma temperature ($n=0$ Fourier component) as functions of time for various fragment numbers.}
	\label{fig:10}
\end{figure}

The SPI process is then simulated for the same Ne impurity injection level with various plume shapes and fragment fineness. In a simulation, all the fragments have the same radius. The results of these cases are presented in Figs. \ref{fig:8}, \ref{fig:9}, and \ref{fig:10}. As fragments become finer, ablated impurities increase sharply, depositing more impurity in the plasma and cooling it more effectively. In tokamaks with higher plasma energy, finer fragments may lead to a complete ablation of impurities near the boundary, as often observed in MGI.

\subsection{Dependence on injection level}
\begin{figure}[htbp]
	\centering
	\includegraphics[width=\linewidth]{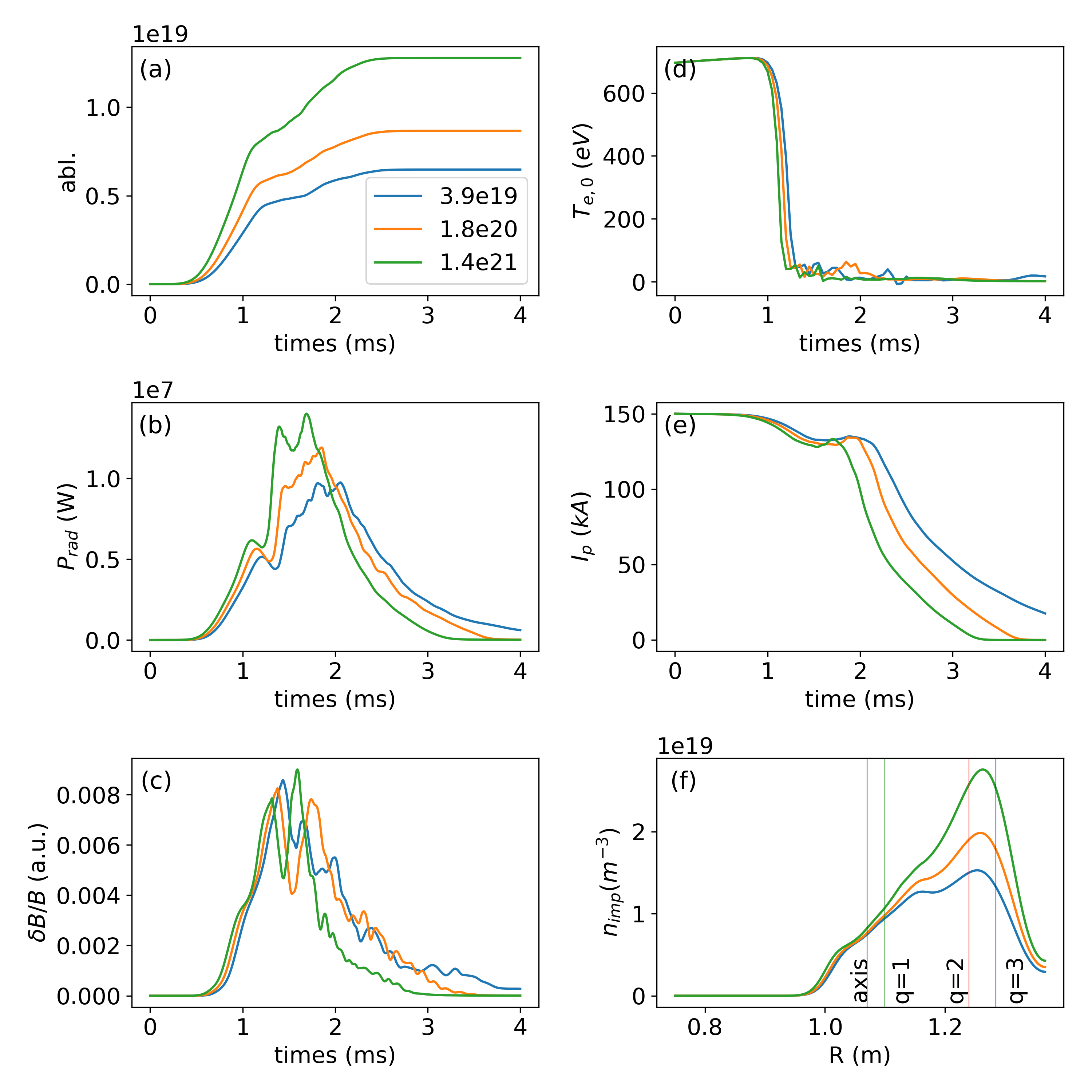}
	\caption{(a) Impurity ablation, (b) radiation power, (c) normalized magnetic energy of $n=1$ toroidal component $\delta B/B = \sqrt{\left ( W_{mag,n=1} / W_{mag,n=0} \right ) }$, (d) core plasma temperatures at $\phi=0$, and (e) plasma current as functions of time, and (f) impurity density profiles immediately after TQ as functions of major radius, for various total impurity injection levels. The black, green, red, and blue solid vertical lines represent the locations of axis, $q=1$, $q=2$, and $q=3$ rational surfaces respectively.}
	\label{fig:11}
\end{figure}

The total injection level or quantity can be varied in several ways, however each of which is associated with another parameter effect in addition to the total injection itself. Here we modulate the total amount of injected impurities by altering the fragment radius. Fig. \ref{fig:11} compares the temporal evolution of impurity ablation, total radiated power, normalized $n=1$ mode perturbed magnetic energy, core temperature, and plasma current, as well as the post-TQ impurity density profiles. As anticipated, increasing the impurity injection quantity leads to greater impurity deposition in the plasma, resulting in higher radiation efficiency and consequently reducing the timescales of both TQ and CQ stages.

\subsection{Dependence on impurity composition}
\begin{figure}[htbp]
	\centering
	\includegraphics[width=\linewidth]{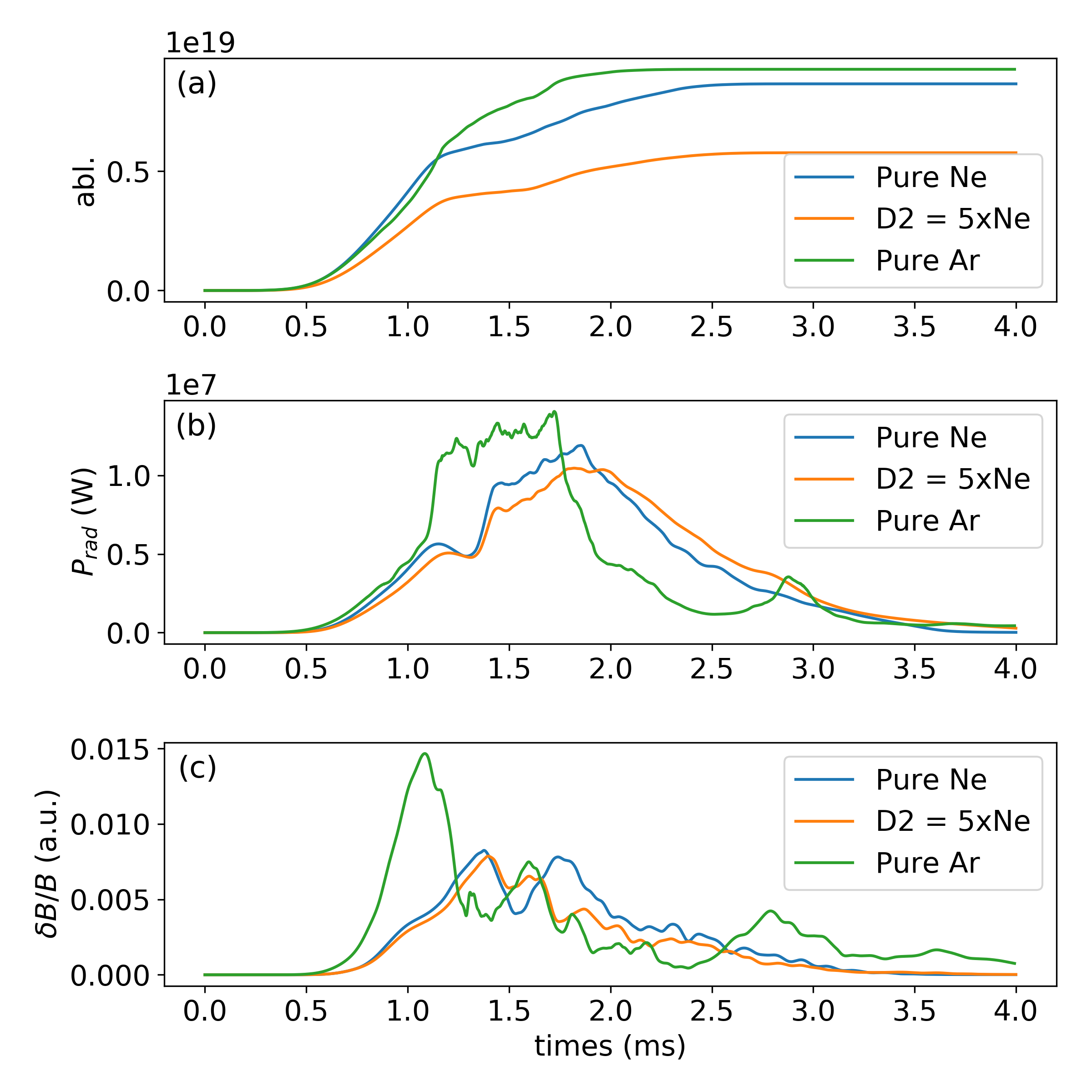}
	\caption{(a) Impurity ablation, (b) radiation power, and (c) normalized magnetic energy of $n=1$ toroidal component $\delta B/B = \sqrt{\left ( W_{mag,n=1} / W_{mag,n=0} \right ) }$ as functions of time for various pellet impurity compositions.}
	\label{fig:12}
\end{figure}

\begin{figure}[htbp]
	\centering
	\includegraphics[width=\linewidth]{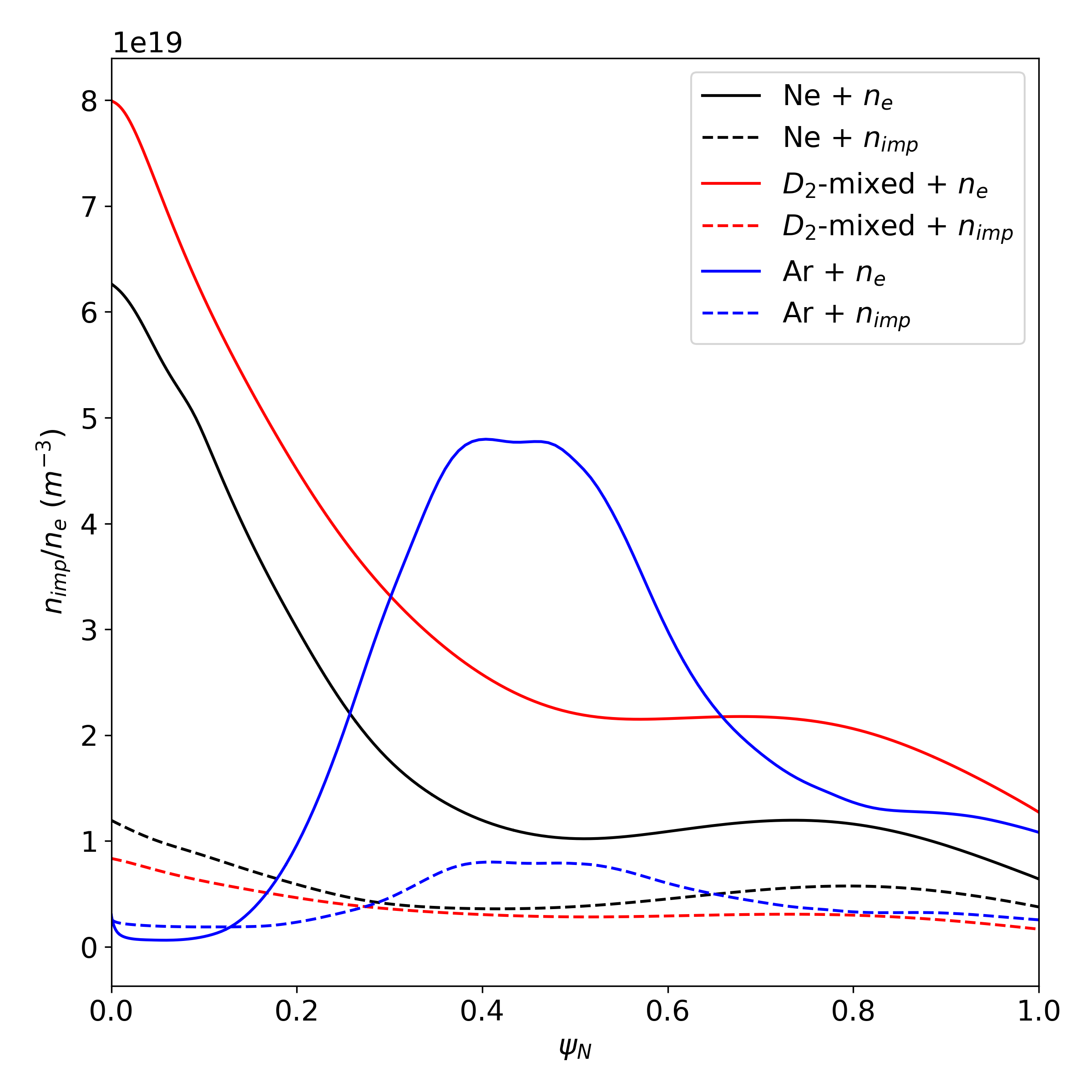}
	\caption{Impurity density and increased electron density profiles as functions of initial normalized poloidal magnetic flux after TQ ($t=1.3$ ms) for various pellet impurity compositions.}
	\label{fig:13}
\end{figure}

The impurity composition of pure $Ne$ particles, mixed $Ne$ and $D_{2}$ particles, and pure $Ar$ particles are compared, which are found to have different ablation characteristics and radiation cooling intensity during flight. Moreover, the $D_{2}$ mixed in pellet can enhance the local pressure. In most experimental studies on deuterium-neon mixed pellet SPI, the neon mixing fraction is generally kept low, resulting in significantly reduced total radiative power compared to the pure-neon SPI. In the present simulation, a high-neon-fraction mixed pellet is employed, which substantially increases the electron density while maintaining radiative cooling levels comparable to those of the pure-neon SPI. For the $Ne$ pellet doped with 5 times molar $D_{2}$, its ablation is less than that of pure $Ne$ pellet, whereas the radiation and the magnetic perturbation are of almost the same level [Fig. \ref{fig:12}]. Although the impurity density of $Ne$ in the case of mixed $D_{2}$ pellet is not as high as that of the $Ne$ pure pellet, the increased electron number density is much higher [Fig. \ref{fig:13}]. Comparing the SPI simulation results between the $Ar$ and $Ne$ pure pellet, the $Ar$ pellet produces a higher radiation power [Fig. \ref{fig:12}(b)],  but its penetration  is less deep to the core region [Fig. \ref{fig:13}].

\subsection{Dependence on injection location and multiple injectors}
\begin{figure}[htbp]
	\centering
	\includegraphics[width=0.8\linewidth]{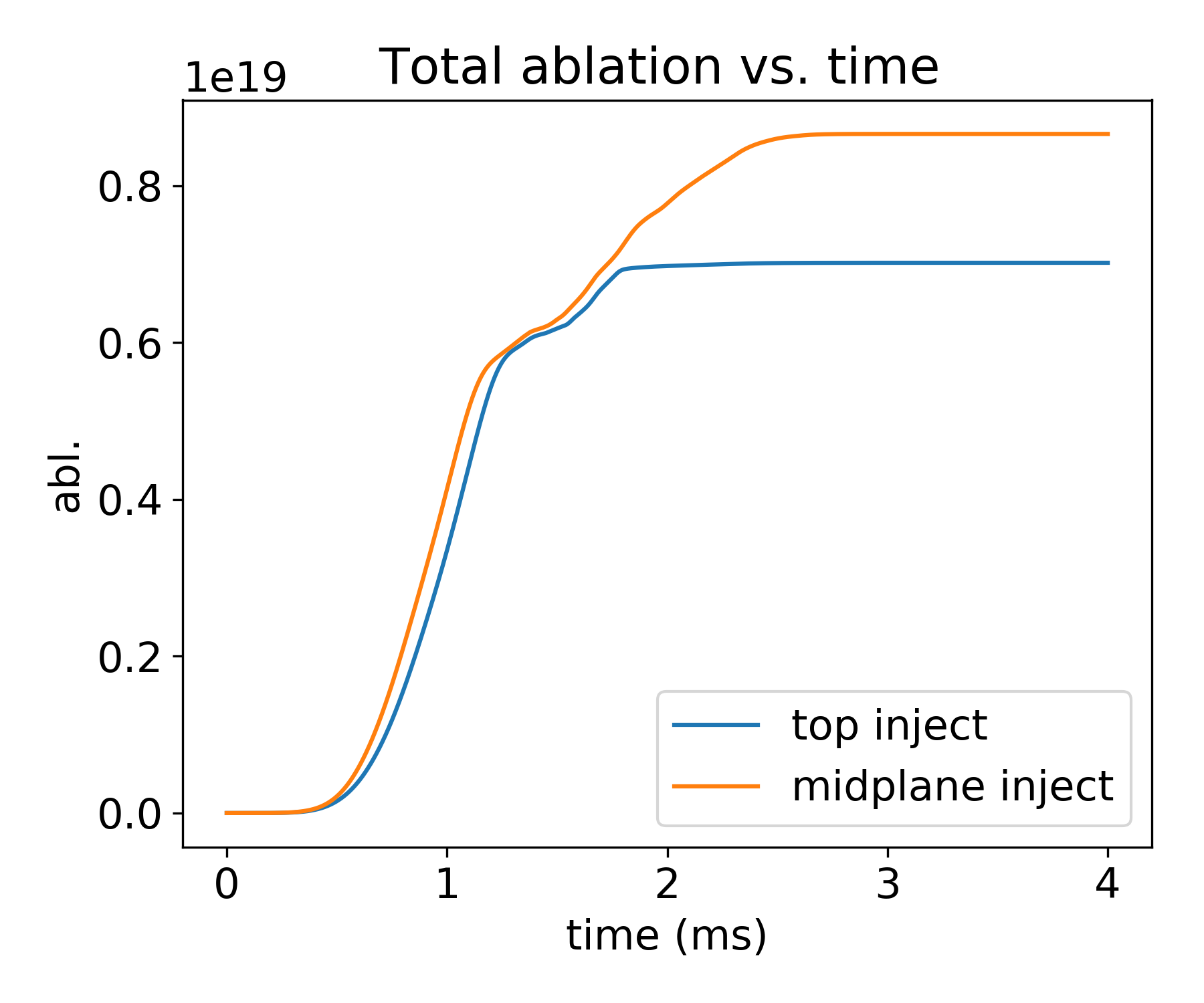}
	\caption{Total impurity ablations as functions of time for top and midplane injections.}
	\label{fig:14}
\end{figure}

\begin{figure}[htbp]
	\centering
	\includegraphics[width=0.8\linewidth]{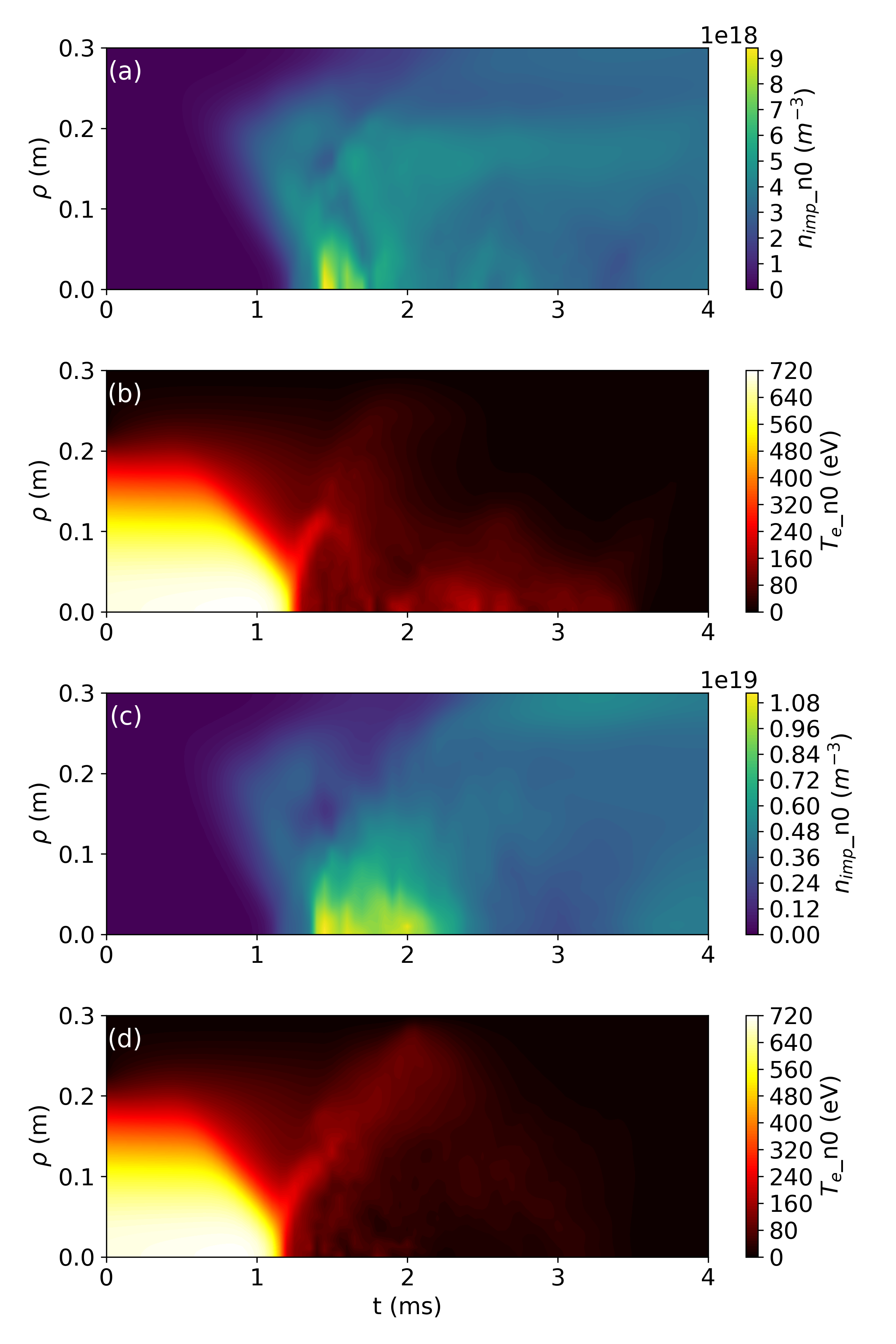}
	\caption{Radial distributions of impurity density (a and c) and plasma temperature (b and d) ($n=0$ Fourier component) as functions of time for top (a and b) and midplane injections (c and d).}
	\label{fig:15}
\end{figure}

\begin{figure}[htbp]
	\centering
	\includegraphics[width=\linewidth]{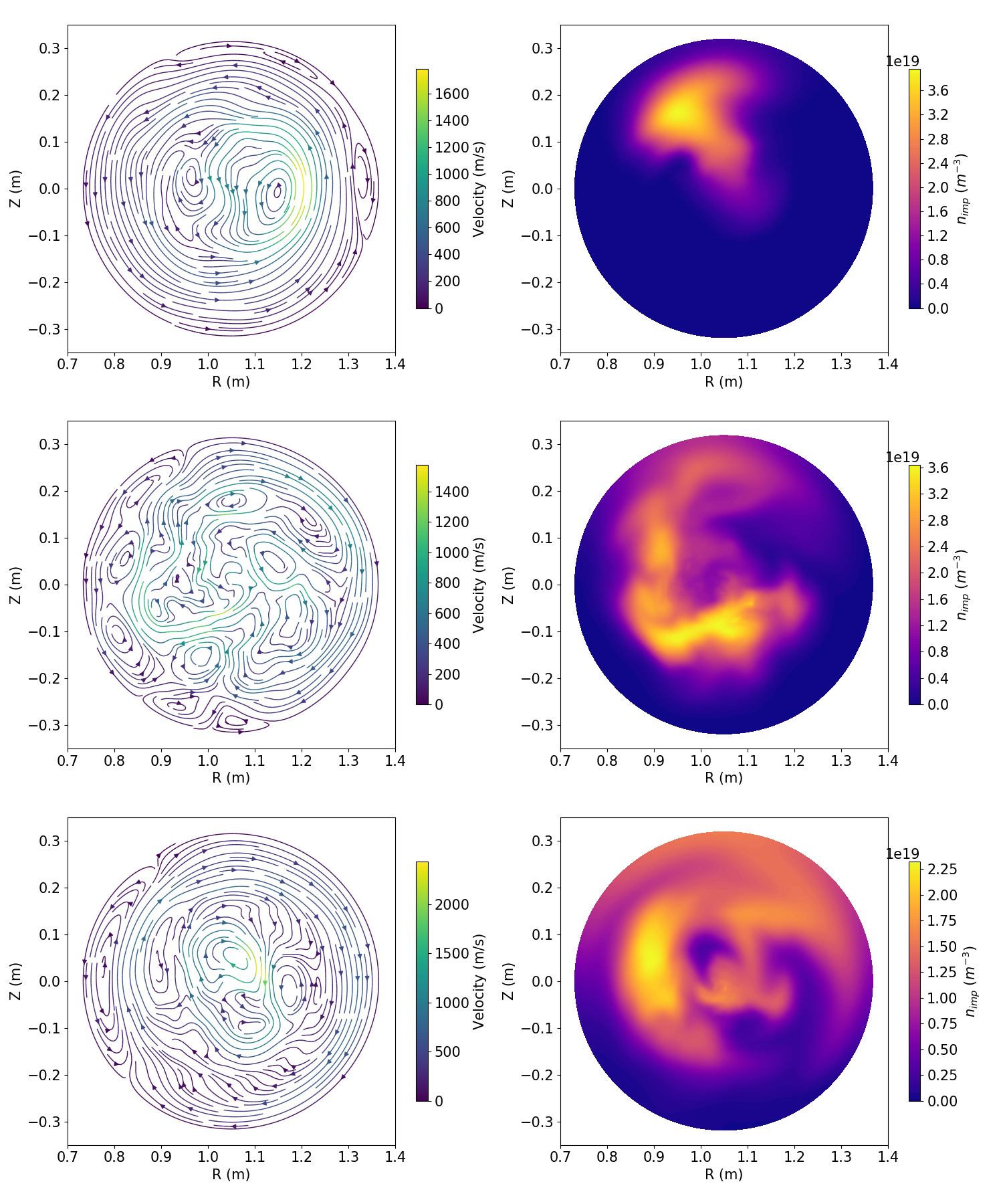}
	\caption{Poloidal flow streamlines (left column) and impurity density distribution (right column) on the $ \phi=0 $ poloidal plane at various times for the top injection case. The three rows from top to bottom correspond to t = 1.3 ms, 2 ms, and 2.5 ms, respectively.}
	\label{fig:16}
\end{figure}

\begin{figure}[htbp]
	\centering
	\includegraphics[width=\linewidth]{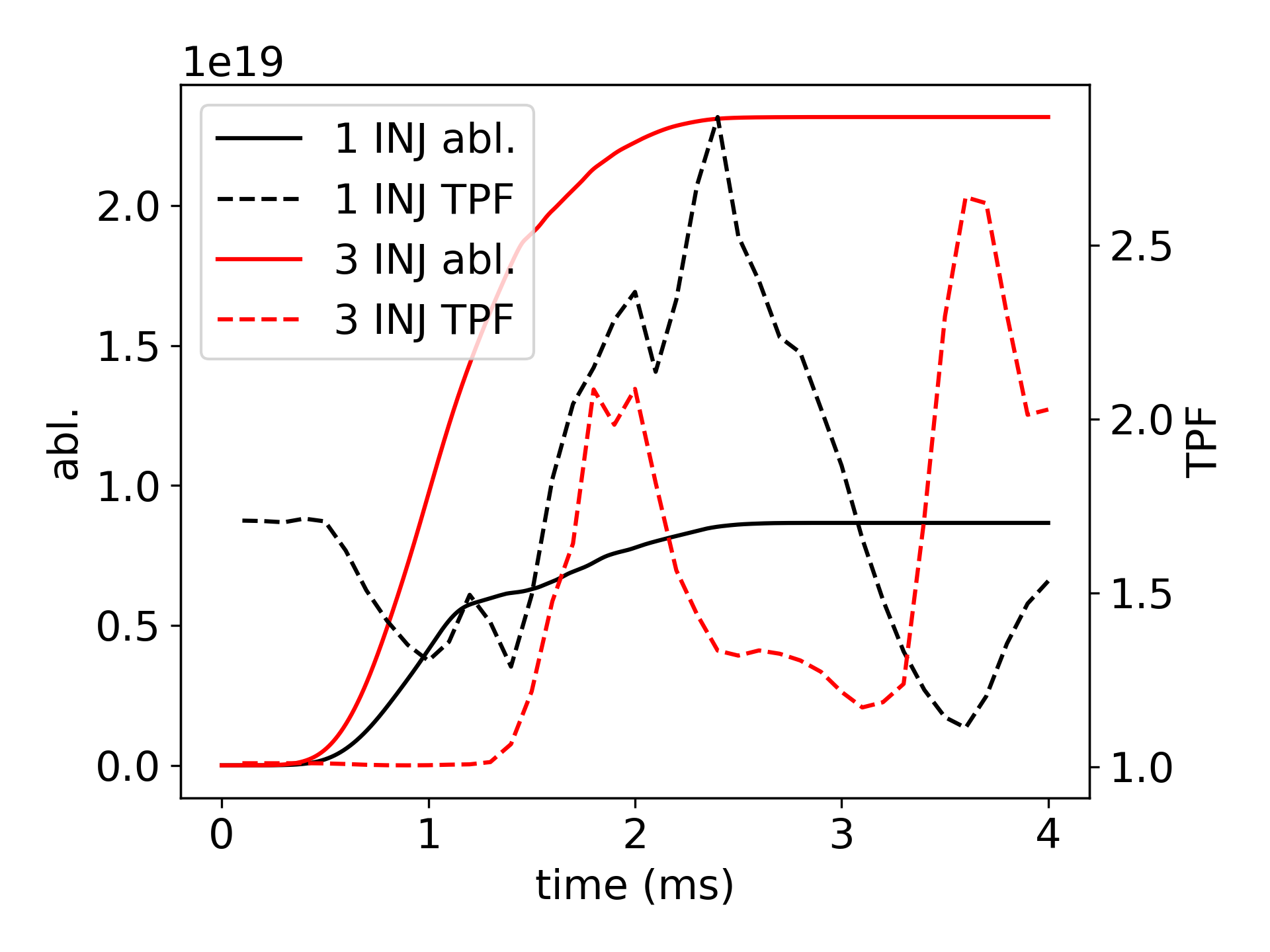}
	\caption{Impurity ablations (solid lines) and radiation Toroidal Peaking Factors (TPFs) (dashed lines) as functions of time for single-pellet (dark lines) and three-pellet injections (red lines).}
	\label{fig:17}
\end{figure}

\begin{figure}[htbp]
	\centering
	\includegraphics[width=\linewidth]{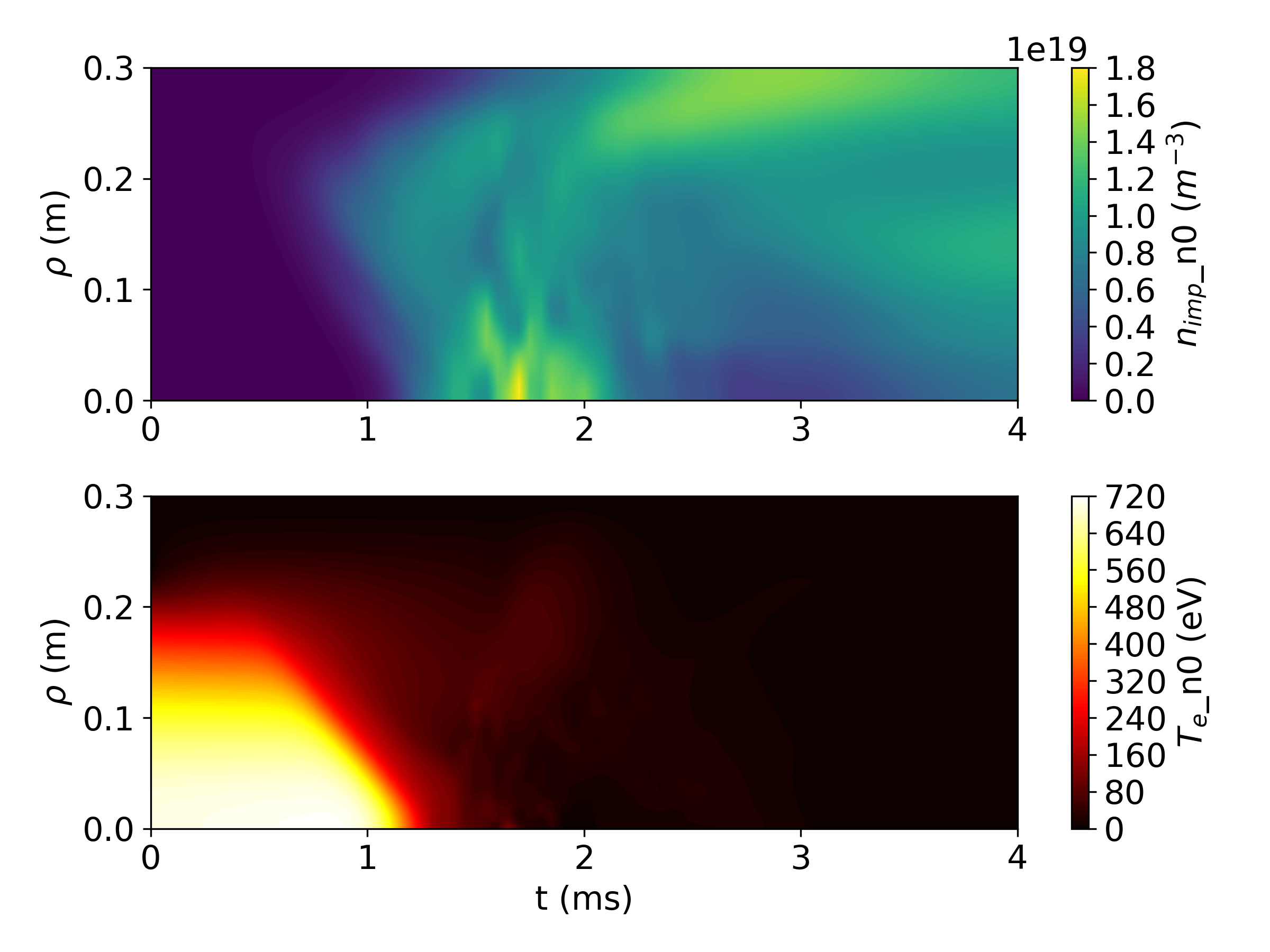}
	\caption{Radial distributions of impurity density (top) and plasma temperature (bottom) ($n=0$ Fourier component) as functions of time for the three-pellet injection case.}
	\label{fig:18}
\end{figure}

Simulation cases with a top injection and three toroidally evenly located injectors in midplane are compared with the baseline case. The poloidal plasma rotation driven by the top injection is shown to reduce both impurity ablation and plasma cooling efficiency [Figs. \ref{fig:14} and \ref{fig:15}]. In the top injection scenario, the amount of ablated impurities after TQ is significantly reduced [Fig. \ref{fig:14}]. The core impurity assimilation density is lower, resulting in weaker core cooling [Fig. \ref{fig:15}]. The plasma first rotates counterclockwise, then decelerates and reverses to clockwise rotation [Fig. \ref{fig:16}]. The direction of plasma rotation is governed by the balance between the torques from the $ J \times B $ force and the thermal pressure gradient\cite{zeng_species_2023}. At t = 2ms, the impurity distribution rotates toward the bottom, coinciding with the approximate location where the pellet fragments arrive. Since the fragments are in a radiatively cooled background plasma, their ablation and assimilation are reduced. For the midplane injection case, although the poloidal flow is also present, it remains symmetric about the midplane and thus does not induce net plasma poloidal rotation. Poloidal flow influences the fragment ablation and assimilation by modifying the background plasma parameters near the fragments, which could benefit the poloidal radiation uniformity. Comparison between the toroidally uniform three-pellet injection and the single-pellet injection shows that the multi-pellet injection enhances the impurity assimilation quantity by nearly a factor of the pellet number, effectively increasing the impurity assimilation density while reducing the radiation TPF [Figs. \ref{fig:17} and \ref{fig:18}], which is a critical engineering design parameter for disruption mitigation systems in devices such as ITER. A lower TPF indicates a more uniform toroidal distribution of radiative power, which helps prevent localized overheating and damage to first-wall materials caused by highly concentrated radiation. A toroidally uniform multi-pellet injection scheme can not only reduce the TPF significantly, but also increase the amount of impurity injection. This demonstrates the potential advantage of such scheme in meeting ITER's requirements for radiation symmetry and massive level of impurity injection.

\subsection{Dependence on plasma resistivity}
\begin{figure}[htbp]
	\centering
	\includegraphics[width=\linewidth]{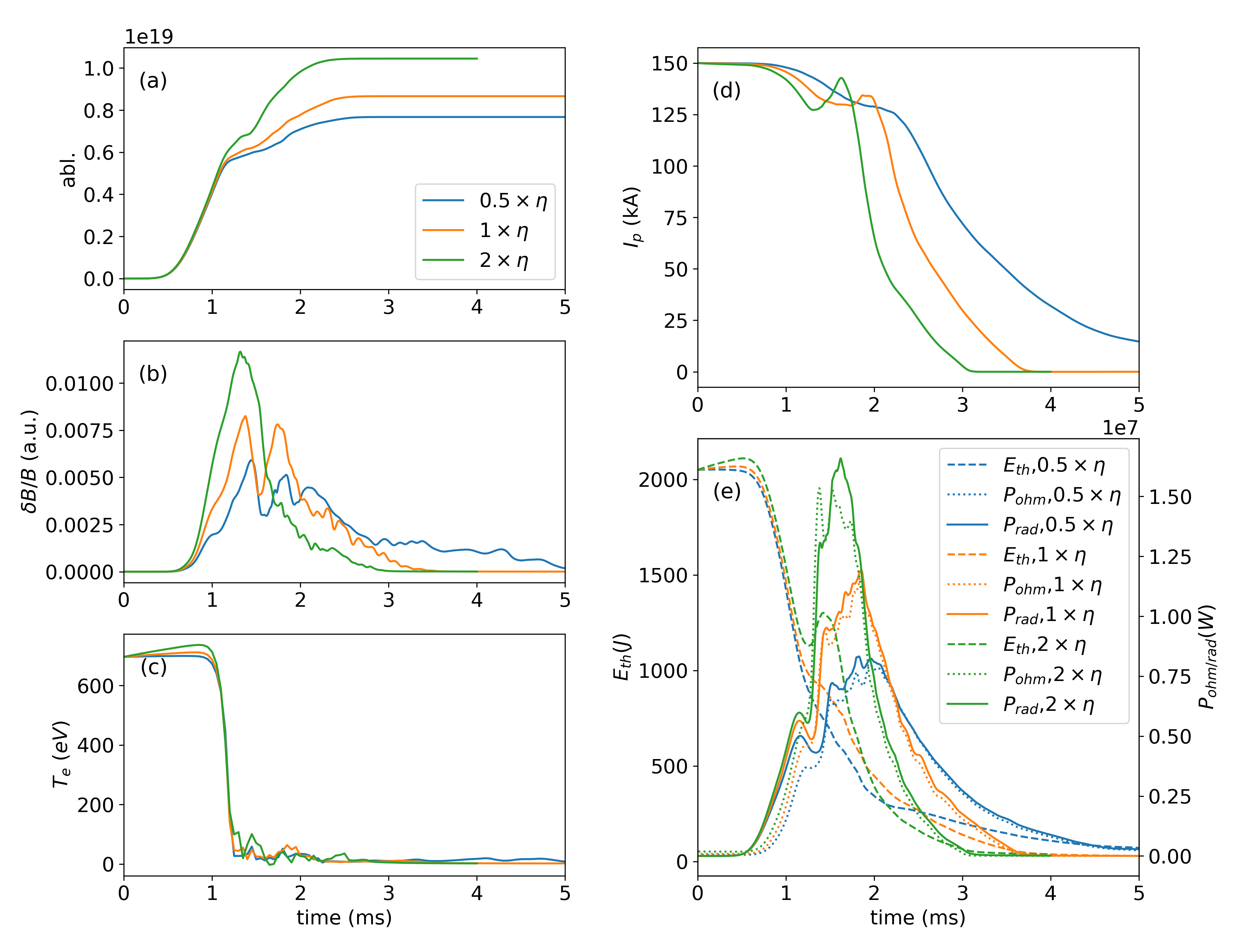}
	\caption{(a) Impurity ablation, (b) normalized magnetic energy of $n=1$ toroidal component $\delta B/B = \sqrt{\left ( W_{mag,n=1} / W_{mag,n=0} \right ) }$, (c) core plasma electron temperature at $\phi=0$, (d) plasma current, and (e) thermal energy, radiation power and Ohmic heating power as functions of time for various plasma resistivity values.}
	\label{fig:19}
\end{figure}

Comparative simulations reveal that the plasma resistivity exhibits negligible influence on either the duration of the TQ phase or the impurity ablation during TQ [Fig. \ref{fig:19}(a) and (c)]. Instead, the impurity ablation during TQ may depend predominantly on plasma internal energy. However, after the TQ phase, higher resistivity leads to increased impurity ablation, higher radiation power, stronger Ohmic heating, more intense magnetic perturbations, and more pronounced current spikes [Fig. \ref{fig:19}(b) and (d)]. Post-TQ impurity ablation shows significant variations across cases. Elevated resistivity leads to stronger Ohmic heating, enabling the system to reach the heating-radiation equilibrium at higher temperatures. With the double resistivity, the Ohmic heating power even temporarily exceeds the total energy loss. This is evidenced by the higher residual core temperature in Fig. \ref{fig:19}(c) and transient internal energy recovery in Fig. \ref{fig:19}(e). The Ohmic heating power does not necessarily become halved or doubled in the half-resistivity and double-resistivity cases. In addition to the minor plasma current variations, this deviation mainly derives from the temperature-dependency of the Spitzer resistivity ($ \eta \propto T^{-3/2} $). The impurity ablation rate ($ \propto T^{5/3} $) intensifies with higher resistivity, delivering enhanced impurity deposition into the plasma. Higher temperature and greater impurity deposition together enhance radiation power. Higher resistivity also leads to greater initial current reduction, stronger current profile contraction, sharper current spikes, and eventually the increased magnetic energy released through magnetic reconnection driven by stronger magnetic perturbations [Fig. \ref{fig:19}(b) and (d)]. In large tokamak devices such as ITER, their inherently low baseline resistivity leads to a significant reduction in ohmic heating power after TQ, accompanied by relatively limited impurity ablation. The substantial portion of the stored magnetic energy is converted into the energy of the REs, instead of being dissipated resistively.

\subsection{Dependence on thermal conductivity}
In previous simulations, we utilized a parallel thermal conductivity with $ \kappa_{\parallel0}=10^6 $. Here, simulations with various parallel thermal conductivity coefficient $ \kappa_{\parallel0} $ are compared to determine its impact on the impurity assimilation rate and the radiation intensity, among others. In addition, we also consider a case where the parallel thermal conductivity and perpendicular thermal conductivity are constants ($ \kappa_{\parallel}=10^6 , \kappa_{\perp}=1$), where the thermal conductivity is uniform and does not decrease with the collapse of the temperature profile.

\begin{figure}[htbp]
	\centering
	\includegraphics[width=\linewidth]{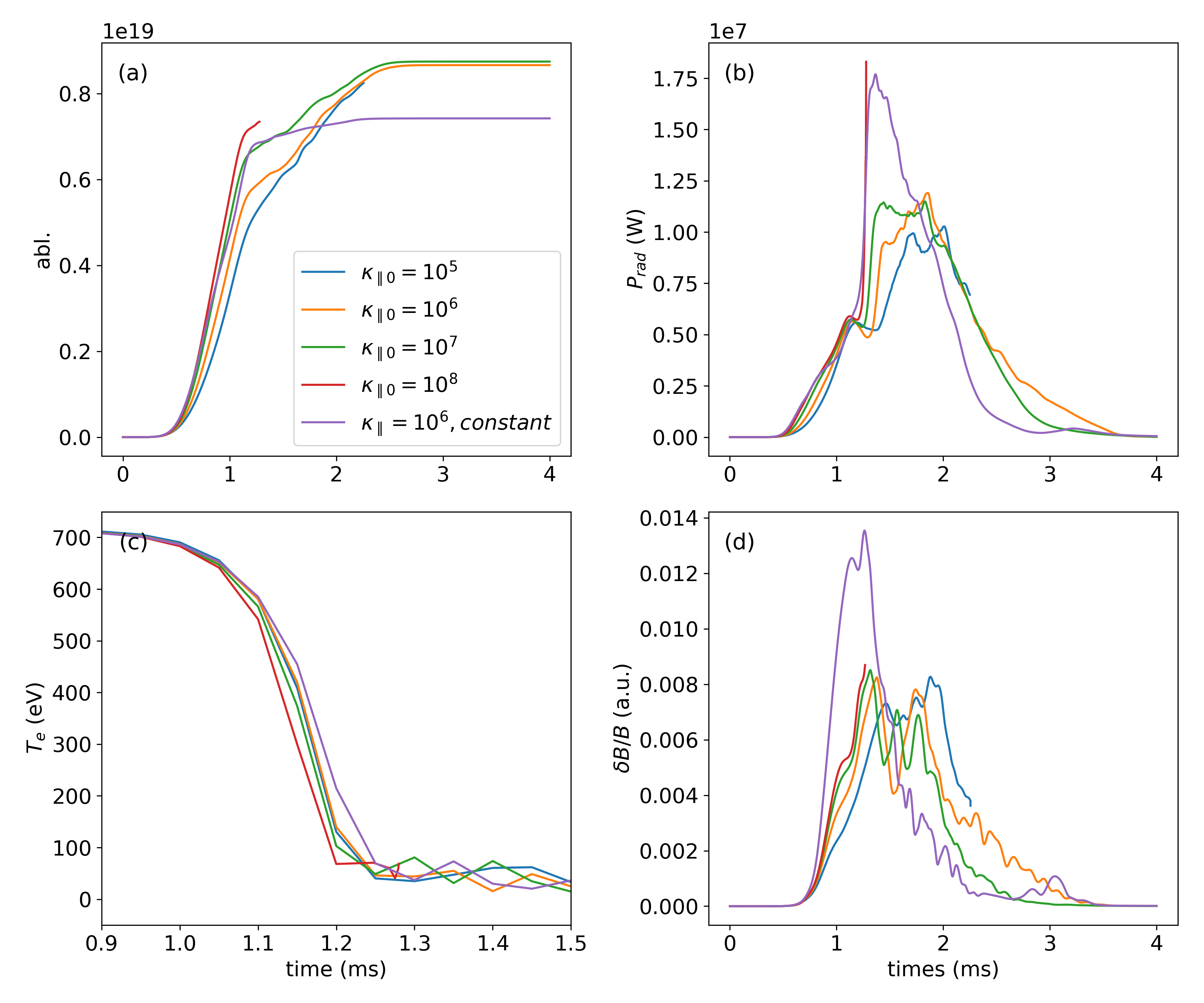}
	\caption{(a) Impurity ablation, (b) radiation power, (c) core plasma electron temperature at $\phi=0$, and (d) normalized magnetic energy of $n=1$ toroidal component $\delta B/B = \sqrt{\left ( W_{mag,n=1} / W_{mag,n=0} \right ) }$ as functions of time for various parallel thermal conductivity coefficients.}
	\label{fig:20}
\end{figure}

\begin{figure}[htbp]
	\centering
	\includegraphics[width=\linewidth]{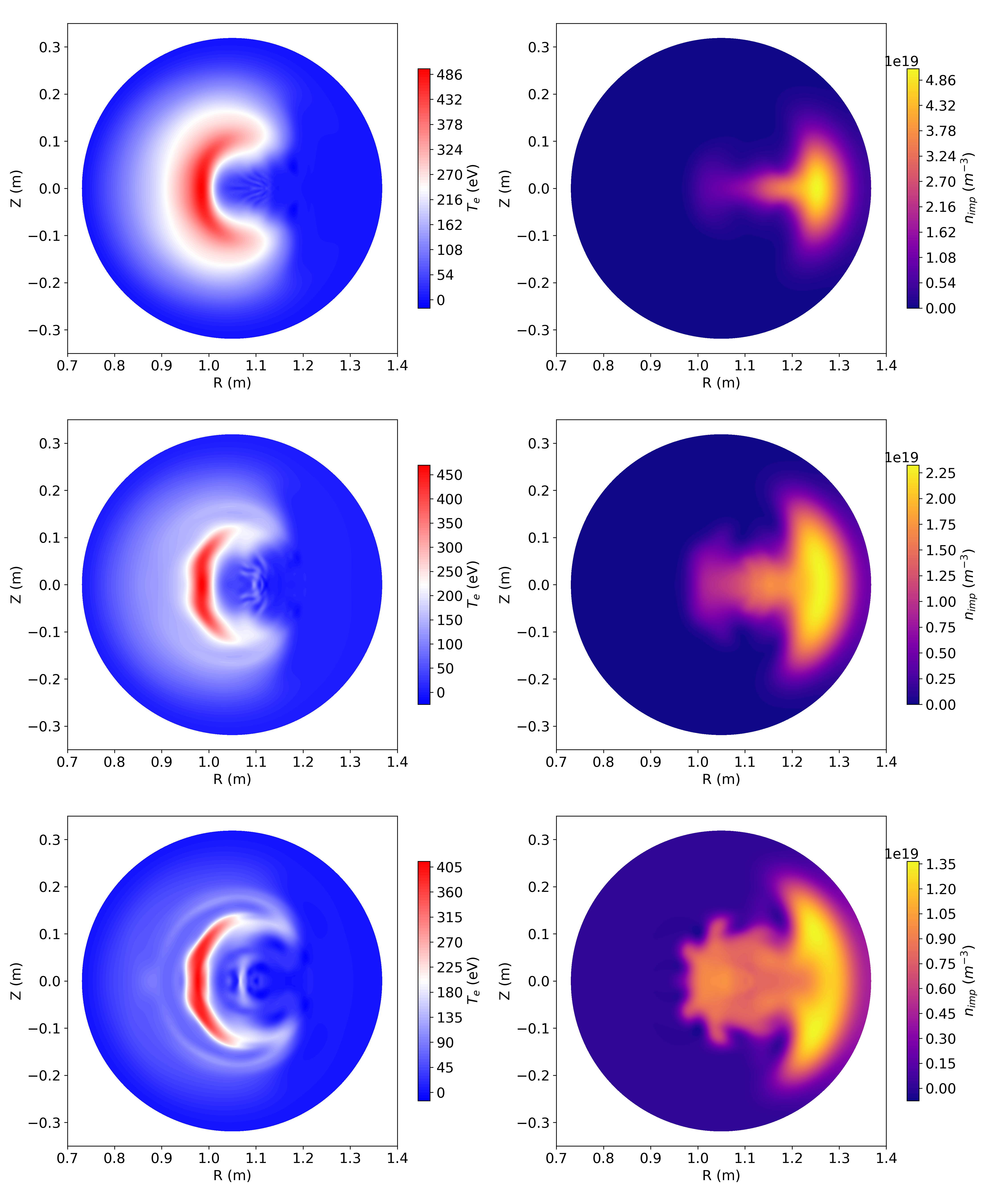}
	\caption{Contours of electron temperature (left) and impurity density (right) in poloidal cross-section ($\phi = 0$) at time $t=1.3$ ms for parallel thermal conductivity coefficients: $ \kappa_{\parallel0}=10^5 $ (top), $ \kappa_{\parallel0}=10^6 $ (middle), and $ \kappa_{\parallel0}=10^7 $ (bottom).}
	\label{fig:21}
\end{figure}

\begin{figure}[htbp]
	\centering
	\includegraphics[width=\linewidth]{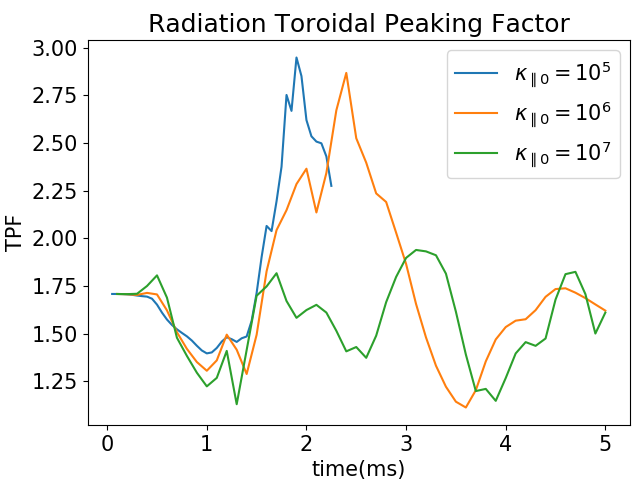}
	\caption{Radiation Toroidal Peaking Factors (TPFs) as functions of time for various parallel thermal conductivity coefficients.}
	\label{fig:22}
\end{figure}

Fig. \ref{fig:20} indicates that increasing the parallel thermal conductivity enhances the impurity ablation, radiation intensity, and magnetic perturbation intensity in the early stage (before and during TQ), but slightly decreases the duration of the TQ by less than 0.1 ms [Fig. \ref{fig:20}(c)]. The impurity distribution and temperature contours during TQ process ($t=1.3$ ms) show that with higher parallel thermal conductivity, impurities cool the magnetic surface more thoroughly as fragments travel from the edge to the core, resulting in minimal temperature differences between the high and low field sides of the magnetic surface [Fig. \ref{fig:21}]. There is a significant difference in the distribution of impurity deposition among the three cases. Whereas the macroscopic plasma responses are similar, the substantial differences in the processes of impurity penetration and temperature profile evolution may affect the depth of impurity deposition and the increase in core electron density, which could be relevant when considering the need to raise electron density to suppress runaway electrons. As the parallel thermal conductivity increases, the TPF values tend to decrease [Fig. \ref{fig:22}].

\section{Conclusion and discussion}
The impurity assimilation efficiency during an SPI-induced disruption process in a J-TEXT-like medium-sized tokamak has been studied using NIMROD simulations. The simulation results indicate that a slower fragment flight speed can improve the impurity assimilation and enhance MHD activity, but it does not benefit the penetration of impurities inward, whereas finer fragments significantly increase the ablation efficiency and the cooling performance. The slower fragment velocity leads to higher impurity assimilation but poorer core penetration depth. This finding suggests a potential trade-off in SPI optimization. In scenarios aimed at suppressing REs, achieving a sufficiently high core electron density is crucial for RE mitigation. Consequently, an optimal fragment velocity window exists, which requires balancing high assimilation efficiency and adequate core impurity deposition. Future research should incorporate runaway electron generation models to determine this optimal velocity for specific devices and disruption scenarios.

The $D_{2}$--mixed pellet is more effective in increasing electron number density, and has a lower impurity assimilation than the pure $Ne$ pellet, but they are almost identical in terms of the induced radiative cooling level. Top injection drives poloidal rotation that reduces ablation and cooling efficiency, whereas midplane injection avoids net rotation and maintains symmetric dynamics. Toroidally uniform multi-pellet injection enhances impurity assimilation by nearly the number of pellets and reduces radiation TPF, which is thus able to meet the challenges of radiation asymmetry and the need for large impurity injection in large tokamaks. Plasma resistivity has a negligible impact on the TQ duration and the impurity ablation, which depend primarily on plasma internal energy. However, higher resistivity during the post-TQ phase enhances the Ohmic heating, impurity ablation, radiation power, magnetic perturbations, and current spikes, due to the temperature-dependent resistivity scaling. The evolution of magnetic field and current distribution is closely related to the resistivity variation. In addition, higher parallel thermal conductivity significantly enhances impurity transport and radiative cooling efficiency in the parallel direction.

Nevertheless, the SPI simulations here have adopted several assumptions. Realistic plume is a mixture of solid, liquid and gas after the pellet shattered by shattering tube. As the pellet velocity increases, the solid content decreases. In simulation here they are assumed as all solid fragments instead. The ablation process of fragments is also simplified. The deposition of neutral impurities in the poloidal plane and toroidal direction is modeled using a Gaussian distribution. In addition, the rocket effects of fragments were not taken into account in this simulation, which may have a significant impact on the inward penetration of impurities in the plasma\cite{kong_interpretative_2024}. The rocket effect can significantly decelerate impurity fragments. Taking this effect into account, the penetration capability of impurities may become even more limited.

Future work will focus on extending this modeling framework to investigate rapid shutdown scenarios in larger tokamak configurations with higher plasma temperatures and energies, particularly relevant for larger sized and ITER-scale devices. Moreover, to achieve higher electron density before or during TQ for suppressing runaway electron generation, various schemes need to be simulated for validation, such as the 2-stage injection\cite{vallhagen_effect_2022}, i.e. a pure $ D_{2}$ SPI followed by a mixed $ D_{2}$-impurity SPI. In addition, modeling efforts are planned to improve the physical fidelity of the SPI model implemented in the NIMROD code.

\begin{acknowledgments}
We are grateful for the supports from the NIMROD team and the J-TEXT team. This work was supported by the National MCF Energy R\&D Program of China under Grant No. 2019YFE03050004, the U.S. Department of Energy Grant No. DE-FG02-86ER53218, and the Hubei International Science and Technology Cooperation Project under Grant No.2022EHB003. The computing work in this paper was supported by the Public Service Platform of High Performance Computing by Network and Computing Center of HUST.
\end{acknowledgments}

\section*{Data Availability Statement}
The data that support the findings of this study are available from the corresponding author upon reasonable request.

\nocite{*}
\bibliography{references}

\newpage

\end{document}